\documentclass[lettersize,journal]{IEEEtran}
\usepackage{amsmath,amsfonts,amssymb,bm}
\usepackage{algorithmic}
\usepackage{algorithm}
\usepackage{array}
\usepackage{textcomp}
\usepackage{stfloats}
\usepackage{url}
\usepackage{verbatim}
\usepackage{graphicx}
\usepackage[caption=false,font=footnotesize]{subfig}
\usepackage{cite}
\usepackage{color}
\usepackage{multirow}
\usepackage{todonotes}
\usepackage{booktabs} 
\usepackage[flushleft]{threeparttable}
\usepackage{balance}
\usepackage{framed}
\usepackage{xspace} 
\usepackage{makecell}
\newcolumntype{C}[1]{>{\centering\arraybackslash}m{#1}}
\usepackage{pifont}
\newcommand{\step}[1]{\raisebox{-0.1ex}{\large\ding{\numexpr171+#1\relax}}}
\usepackage[most]{tcolorbox}

\newcommand{\toolname}{TraceMsg\xspace}
\newtcolorbox{rqsummary}{
    enhanced,
    colback=gray!8,
    colframe=gray!50,
    boxrule=0.5pt,
    arc=2mm,
    left=2mm,
    right=2mm,
    top=1mm,
    bottom=1mm
}

\begin{document}

\title{Beyond Similarity-Based Generation: Guiding Commit Message Generation with Issue–Commit Traceability}

\author{
\IEEEauthorblockN{
Chunying Zhou\IEEEauthorrefmark{1}\IEEEauthorrefmark{4},
Gong Chen\IEEEauthorrefmark{1}\IEEEauthorrefmark{4},
Xiaoyuan Xie\IEEEauthorrefmark{1}\IEEEauthorrefmark{5},
Yi Song\IEEEauthorrefmark{2}\IEEEauthorrefmark{5},
Yuxuan Yang\IEEEauthorrefmark{1},
Peng He\IEEEauthorrefmark{3}
}

\IEEEauthorblockA{\IEEEauthorrefmark{1}School of Computer Science, Wuhan University, Wuhan, China}

\IEEEauthorblockA{\IEEEauthorrefmark{2}School of Cyber Science and Engineering, Wuhan University, Wuhan, China}

\IEEEauthorblockA{\IEEEauthorrefmark{3}School of Cyber Science and Technology, Hubei University, Wuhan, China}

\IEEEauthorblockA{\textbraceleft zcy9838, chengongcg, xxie, yisong, yxyang3\textbraceright@whu.edu.cn; penghe@hubu.edu.cn}

\thanks{\IEEEauthorrefmark{4}Chunying Zhou and Gong Chen contributed equally to this work.}
\thanks{\IEEEauthorrefmark{5}Xiaoyuan Xie and Yi Song are the co-corresponding authors.}
}

\maketitle

\begin{abstract}
Commit messages provide records of code changes and help developers understand and maintain software. To reduce the manual effort involved in writing them, commit message generation (CMG) approaches automatically generate commit messages that describe code changes in natural language.
Existing CMG approaches generally treat CMG as a standalone task and optimize generated messages primarily for similarity to developer-written references. 
However, this objective may not properly reflect the practical usefulness of generated commit messages. This limitation matters in current AI-assisted software engineering workflows, where commit messages are used as shared information to serve different software engineering activities.
To fill this gap, we propose \toolname, which uses issue–commit \underline{trace}ability, a representative software engineering activity, to guide commit \underline{m}e\underline{s}sa\underline{g}e generation. \toolname is a collaborative multi-agent framework that coordinates message generation, traceability reasoning, and feedback-guided refinement, allowing generated messages to serve traceability while using reasoning feedback to guide issue-conditioned refinement.
We evaluate \toolname on six open-source projects. The experimental results show that \toolname achieves an average Hit@1 of 0.893 and an MRR of 0.911, improving Hit@1 and MRR by 13.16\% and 38.75\%, respectively, over the best baseline for each metric.
\end{abstract}


\section{Introduction}
\IEEEPARstart{C}{ommit} messages are important software artifacts that record the evolution of software systems and support various software engineering activities, including code review~\cite{badampudi2023modern}, program comprehension~\cite{fang2024large}, software maintenance~\cite{ren2023deeparc}, and software evolution~\cite{zhang2026survey}. To reduce the manual effort of writing commit messages, extensive research has explored automatic commit message generation (CMG). Earlier CMG approaches derived commit messages from patterns learned from historical code changes and developer-written messages using task-specific retrieval or supervised models~\cite{liu2018neural,hoang2020cc2vec,jiang2017automatically,liu2019generating,dong2022fira}. Recently, large language models (LLMs) have empowered CMG to directly interpret code changes and generate natural-language messages through prompting, reasoning, in-context learning, and retrieval augmentation~\cite{li2024only,xue2024automated,xiong2026coracmg,zhang2024automatic,wu2025empirical}. These advances have substantially improved the fluency, specificity, and semantic quality of generated commit messages.

However, existing CMG approaches generally restrict CMG to a standalone task, with improvements primarily demonstrated through higher lexical and semantic similarity to developer-written references. \textbf{Such an optimization objective may not properly reflect the practical usefulness of generated messages across software engineering activities.} In practice, commit messages not only record individual code changes but also carry information that supports communication across different activities.
For example, software maintenance often requires recovering missing links between issues and their fixing commits, which facilitates subsequent fix tracing. In this software activity, commit messages describe code changes in natural language, thereby providing textual evidence for matching issue reports with their commits. However, existing CMG approaches focus primarily on improving similarity to developer-written references. Such similarity captures textual correspondence but does not indicate whether a generated message provides the semantic connection between an issue report and its corresponding code change. 
Consequently, optimizing similarity scores alone may not be sufficient for generating messages that effectively serve the software engineering activities in which they are used.

More importantly, as LLM-based agents take on more software engineering activities previously performed by developers, commit messages are more frequently used to communicate information among agents within automated workflows. \textbf{The ability of commit messages to serve different activities is therefore essential in current AI-assisted software engineering.} Accordingly, CMG should no longer be treated as a static and isolated task, but as a dynamic process designed to serve various activities within automated software engineering workflows.


To address this limitation, we propose \toolname, a collaborative multi-agent framework that treats CMG as a process guided by a particular software engineering activity. 
Among common software engineering activities, we focus on issue--commit traceability because it is an important and representative software maintenance activity that supports maintenance history reconstruction, program understanding, and software evolution~\cite{Akhavan2026LinkAnchor,Huang2026Back,zhang2023ealink,dong2022semi,lin2021traceability}. In AI-assisted software engineering, reliable issue--commit links can further provide LLM-based agents with historical problem--fix information for locating prior fixes, reasoning about repair strategies, and validating code changes. This makes issue--commit traceability a suitable activity for integrating commit message generation with practical use.
Instead of treating CMG as an isolated text generation task, 
\toolname formulates it as a dynamic process of message generation, traceability reasoning, and feedback-guided refinement.
Specifically, a \emph{Generation Agent} analyzes commit diffs and produces structured commit messages containing both implementation changes and inferred development intent. 
These generated messages are then used for issue–commit traceability reasoning by a \emph{Traceability Agent}.
Based on the traceability results, a \emph{Feedback Agent} identifies missing or ambiguous intent information and provides feedback to refine the generated messages. Through this dynamic process, \toolname treats generated commit messages as shared information for issue--commit traceability, using traceability outcomes to assess their practical usefulness and guide their issue-conditioned refinement.

The main contributions of this paper are summarized as follows:

\begin{itemize}

\item \textbf{A practical usability perspective.}
We propose a new perspective in which CMG is guided by how effectively generated messages serve software engineering activities rather than by their static similarity to developer-written references. This perspective reflects the practical usefulness of generated messages across these activities.

\item \textbf{A collaborative multi-agent framework.}
We propose \toolname, a collaborative multi-agent framework that enables CMG to serve a particular software engineering activity in AI-assisted workflows.
\toolname uses issue--commit traceability as a representative activity and coordinates three agents to form a dynamic process of message generation, traceability reasoning, and feedback-guided refinement.

\item \textbf{A comprehensive evaluation.}
We conduct extensive experiments on six open-source projects to evaluate the effectiveness of \toolname and investigate the practical usefulness of generated commit messages in issue--commit traceability.
\end{itemize}

The remainder of this paper is organized as follows. Section~\ref{sec:preliminaries} introduces CMG and its evaluation, presents a motivating example, and outlines our solution strategies. Section~\ref{sec:methodology} presents the overall framework of \toolname. Section~\ref{sec:experimental} describes the experimental setup, followed by the experimental results and analysis in Section~\ref{sec:results}. Section~\ref{sec:discussion} provides analyses of parameter sensitivity, backbone robustness, and efficiency. Section~\ref{sec:threats} discusses the threats to validity of this work. Section~\ref{sec:related_work} reviews related work, and Section~\ref{sec:conclusion} concludes the paper.

\section{Preliminaries}\label{sec:preliminaries}

\subsection{Commit Message Generation and Evaluation}

Due to rapid development cycles and workload pressure, manually writing commit messages can be time-consuming. To reduce this effort, CMG approaches aim to automatically generate natural-language descriptions of code changes (i.e., commit diffs). Besides commit diffs, recent approaches may further exploit auxiliary development artifacts, such as pull requests, similar historical commits, and existing commit messages, to improve the generated commit messages~\cite{li2024only}. Regardless of the input sources, the generated commit message is expected to summarize not only \textit{what} was changed, but ideally also \textit{why} the change was necessary~\cite{tian2022makes, li2023commit}.
To evaluate generation quality, existing studies commonly compare generated messages with corresponding developer-written references using lexical and semantic similarity metrics. Lexical similarity is often measured using BLEU~\cite{papineni2002bleu}, METEOR~\cite{banerjee2005meteor}, and ROUGE-L~\cite{lin2004rouge}, while semantic similarity is measured using embedding-based metrics~\cite{zhang2019bertscore}. Under these evaluation settings, higher similarity to developer-written references is generally regarded as better generation quality. 

\subsection{Motivating Example}

\begin{figure*}[!t]
    \centering
    \subfloat[High similarity between the generated commit message and the reference message.
        \label{fig:motivation_example_a}
    ]{\includegraphics[width=0.56\textwidth]{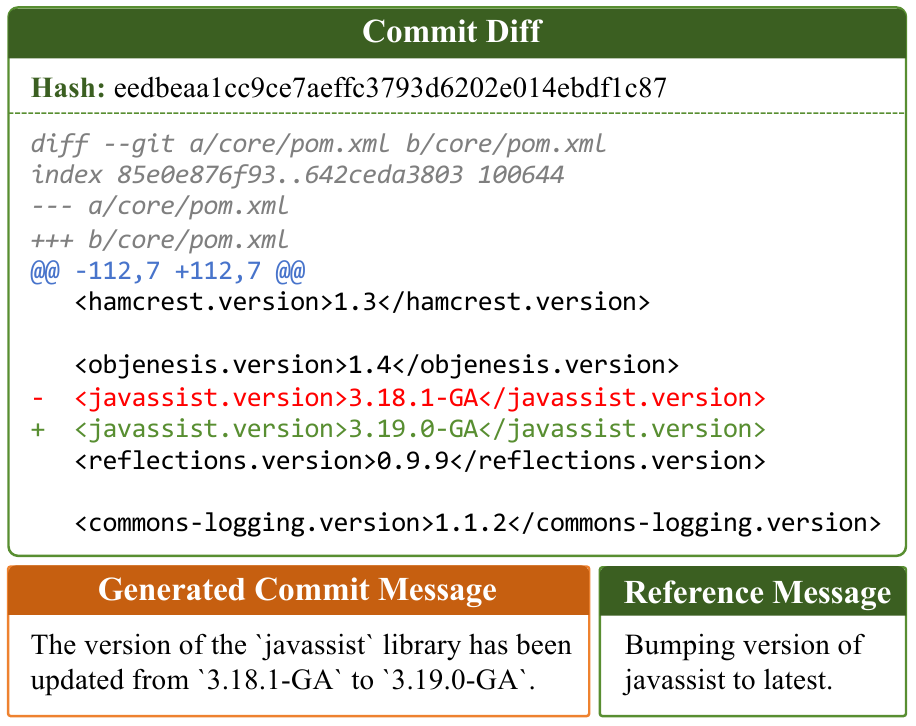}}
    \hfill
    \subfloat[Issue report associated with the commit in Fig.~\ref{fig:motivation_example_a}.
        \label{fig:motivation_example_b}
    ]{\includegraphics[width=0.41\textwidth]{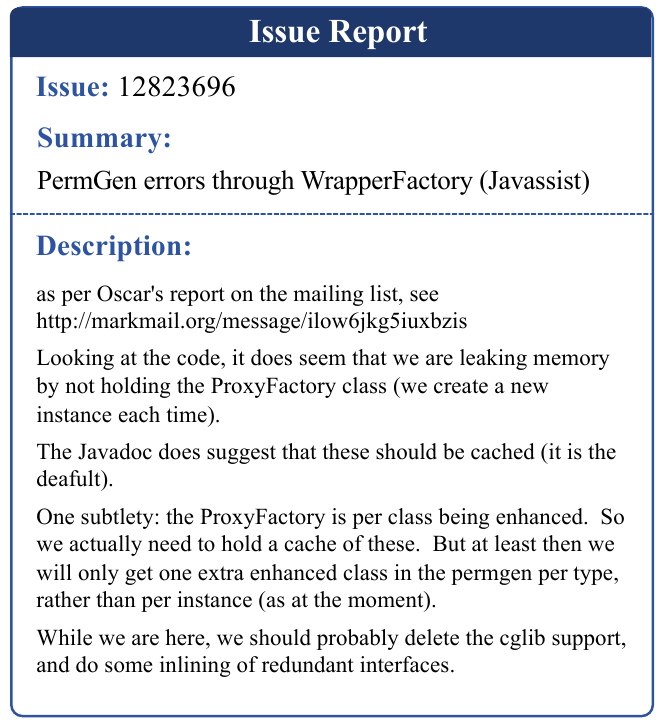}}
    \caption{A motivating example from the \textit{Apache Isis} project illustrating the gap between message generation quality and the usefulness of generated messages in a software engineering activity.}
    \label{fig:motivation_example}
\end{figure*}

To illustrate the gap between message generation quality and the usefulness of generated messages in software engineering activities, we present a motivating example from the \textit{Apache Isis} project in Fig.~\ref{fig:motivation_example}. As shown in Fig.~\ref{fig:motivation_example}(a), we use the state-of-the-art LLM-based CMG setting evaluated by Xue et al.~\cite{xue2024automated}, in which an LLM generates the commit message from the given diff. The generated commit message achieves a ROUGE-L score of 27.27 for lexical similarity and a BERTScore F1 of 85.77 for semantic similarity against the developer-written reference, indicating favorable generation
quality. However, such metrics characterize similarity to the reference without adequately reflecting whether the generated message is useful in software engineering activities. For example, software maintenance often requires associating commits with related issues to facilitate subsequent fix tracing. As shown in Fig.~\ref{fig:motivation_example}(b), the issue concerns \texttt{PermGen} errors caused by repeatedly creating \texttt{ProxyFactory} instances, whereas the generated message only records the \texttt{Javassist} version update without explaining its connection to the reported problem. When a conventional traceability approach~\cite{Huang2026Back} is used to recover this link, the actual fixing commit is ranked only 249th. This example shows that high similarity to a developer-written reference does not necessarily indicate that a generated message can effectively serve the software engineering activities in which it is used.

\subsection{Solution Strategies}

The motivating example shows that the generation quality measured by reference similarity may not reflect whether the generated message can effectively serve the software engineering activity in which it is used. To address this gap, we develop our solution from two aspects. First, we propose a practical usability perspective for CMG. Instead of treating CMG as a static and isolated task, we allow
generated messages to communicate with software engineering activities and use the information revealed during their use to guide further improvement. This transforms CMG from a standalone generation task into a dynamic process in which message generation and practical use mutually inform each other.
Second, we instantiate this idea through issue--commit traceability, an important and representative software maintenance activity~
\cite{Akhavan2026LinkAnchor,Huang2026Back,zhang2023ealink,dong2022semi,lin2021traceability}. Issue reports describe observed failures and maintenance needs at the problem level, whereas commit messages describe the corresponding code changes at the implementation and repair level. The usefulness of a commit message for issue--commit traceability depends on its ability to convey the problem--fix relationship between the reported issue and the corresponding code changes. Therefore, traceability performance directly reflects the practical usefulness of generated messages in this activity, beyond their similarity to developer-written references.
Accordingly, we design \toolname, which assigns message generation,
traceability reasoning, and feedback-guided refinement to specialized agents. Their collaboration allows generated information and reasoning feedback to flow across different activities, making \toolname well suited to multi-agent interaction in AI-assisted software engineering workflows.

\section{Methodology}\label{sec:methodology}

\subsection{Overview}
\begin{figure*}[!t]
  \centering
 \includegraphics[width=\linewidth]{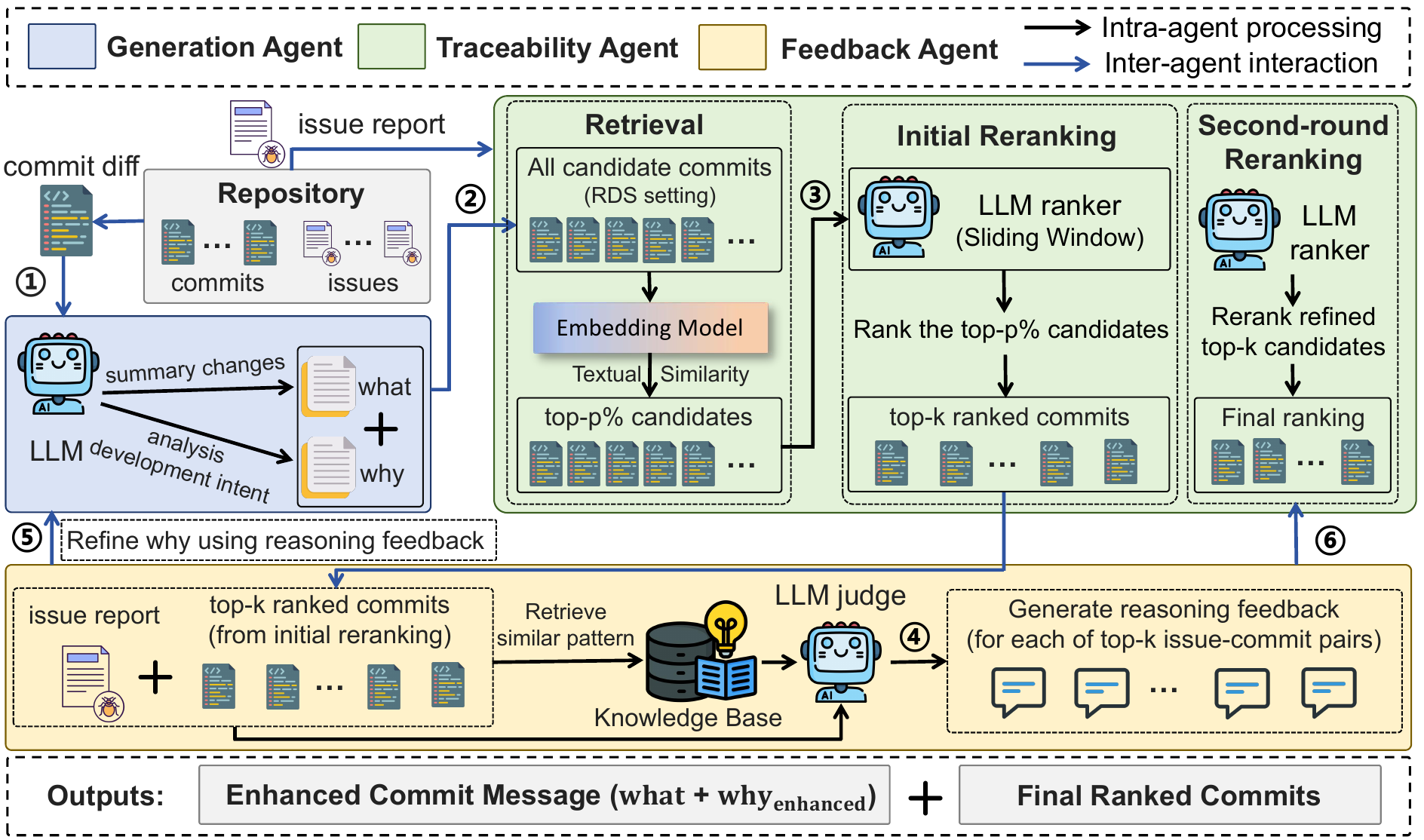}
  \caption{Overall workflow of the proposed framework. \toolname first generates structured commit messages from commit diffs and performs an initial issue--commit traceability process to identify top-$k$ candidates. It then refines the generated \textit{why} using feedback from these candidates and performs a second-round reranking to obtain the final issue--commit ranking.}
  \label{fig:framework}
\end{figure*}

Fig.~\ref{fig:framework} illustrates the overall framework of \toolname, which consists of three collaborative agents: the \textit{Generation Agent}, the \textit{Traceability Agent}, and the \textit{Feedback Agent}. 
Given a repository containing commits and issue reports, the \textit{Generation Agent} first generates a structured commit message for each commit from its commit diff, consisting of a code change summary (\textit{What}) and an inferred rationale of the development intent (\textit{Why}) (step \step{1}). The generated commit messages are then consumed by the \textit{Traceability Agent}, which first retrieves the top-$p\%$ candidate commits (step \step{2}) and performs LLM-based initial reranking to identify the top-$k$ candidate commits for each issue (step \step{3}). 
The given issue and its top-$k$ candidate commits are subsequently analyzed by the \textit{Feedback Agent}. By retrieving similar patterns from a software maintenance knowledge base, the \textit{Feedback Agent} generates reasoning feedback for the top-$k$ issue-commit pairs (step \step{4}). The reasoning feedback is then incorporated by the \textit{Generation Agent} to refine the generated \textit{Why} (step \step{5}), producing enhanced commit messages. Finally, the enhanced commit messages are passed back to the \textit{Traceability Agent} for a second-round reranking over the same top-$k$ candidates (step \step{6}). Consequently, \toolname outputs both the enhanced commit messages and the final ranked commits.

\subsection{Structured Commit Message Generation}\label{sec:CMG}

Given a repository $\mathcal{R}$ containing a set of commits $\mathcal{C}$, the goal of the \textit{Generation Agent} is to generate a structured commit message for each commit $c\in\mathcal{C}$ using only its commit diff. We explicitly decompose the generated commit message into two complementary components: \textit{What}, which summarizes the observable implementation changes, and \textit{Why}, which explains the underlying development intent inferred from the code changes. Formally, the structured commit message associated with commit $c$ is represented as
\begin{equation}
M(c)=\{W(c),\,Y(c)\},
\end{equation}
where $W(c)$ and $Y(c)$ denote the generated \textit{What} and \textit{Why} descriptions, respectively. The generated commit message $M(c)$ serves as the textual information of commit $c$ throughout the subsequent traceability and refinement stages. Since $M(c)$ is generated independently of any issue report, it can be reused across different issue queries.

\paragraph{What Generation}

The \textit{What} component summarizes the implementation-level modifications introduced by a commit. Since commits often involve multiple files with complex dependencies, simply describing individual file changes is insufficient for understanding the overall modification. Therefore, the \textit{Generation Agent} first analyzes each modified file and then summarizes their relationships to produce a description of the code changes. Given the commit diff $D(c)$, the \textit{What} component is generated as
\begin{equation}
W(c)=G_{\mathrm{what}}\!\left(D(c)\right),
\end{equation}
where $G_{\mathrm{what}}(\cdot)$ denotes the LLM-based generation process that produces a structured summary describing both file-level modifications and their inter-file relationships.

\paragraph{Why Inference}

Unlike \textit{What}, which can be directly summarized from observable code changes, the development intent behind a commit is usually implicit and cannot be fully recovered from the commit diff alone. Therefore, instead of directly generating a definitive rationale, the \textit{Generation Agent} infers an initial explanation that captures the most plausible development intent based solely on the available code changes.
To improve the quality of the inferred explanation, the \textit{Generation Agent} first identifies the file that is most relevant to the commit objective according to the generated \textit{What}, and then infers the corresponding development intent conditioned on both the commit diff and the summarized implementation changes. The resulting \textit{Why} is generated as
\begin{equation}
Y(c)=G_{\mathrm{why}}\!\left(D(c),\,W(c)\right),
\end{equation}
where $G_{\mathrm{why}}(\cdot)$ denotes the LLM-based reasoning process that infers the underlying development intent from the code changes. The generated $Y(c)$ represents an initial rationale and will be progressively refined using traceability reasoning feedback in Section~\ref{sec:refinement}.

\subsection{Issue-Commit Traceability}\label{sec:traceability}
Given an issue report $i$ and the structured commit messages generated in Section~\ref{sec:CMG}, the \textit{Traceability Agent} aims to identify the corresponding fixing commit from a large repository of candidate commits. 
Directly ranking all candidate commits using an LLM is computationally prohibitive under the Realistic Distribution Setting (RDS)~\cite{Huang2026Back}, where candidate commits are constructed based on realistic repository activity after issue creation and can reach hundreds or thousands per issue. 
Therefore, \toolname adopts a retrieve-and-rerank strategy to efficiently reduce the candidate space while preserving sufficient candidates for subsequent reasoning.

\paragraph{Candidate Retrieval}
The goal of this stage is to efficiently narrow down the large candidate space while preserving sufficient recall of potentially relevant commits for subsequent LLM-based traceability ranking. For each issue $i$, we denote its candidate commit set as $\mathcal{C}_i^{cand}$. Both issue reports and structured commit messages are encoded into a shared semantic embedding space using the Sentence-Transformers library~\cite{reimers2019sentence} with the all-MiniLM-L6-v2 model. The semantic similarity between issue $i$ and candidate commit $c$ is computed as
\begin{equation}
s(i,c)=\cos\!\left(E(i),E(M(c))\right),
\end{equation}
where $E(\cdot)$ denotes the embedding encoder and $M(c)$ is the structured commit message defined in Section~\ref{sec:CMG}. The candidate commits are ranked according to their similarity scores, and the top-$p\%$ candidates are retained
\begin{equation}
\mathcal{C}_{p\%}(i)= \operatorname{Top}_{p\%} \left(\left\{(c,s(i,c)) \mid c\in\mathcal{C}_i^{cand} \right\}\right).
\end{equation}

\paragraph{Initial Traceability Reranking}

Although dense retrieval effectively filters irrelevant commits, the remaining candidates often exhibit highly similar semantic representations, making them difficult to distinguish using embedding similarity alone. Therefore, we further employ an LLM ranker to perform fine-grained semantic comparison between the issue report and each candidate commit.
We adopt a sliding-window ranking strategy inspired by RankGPT~\cite{sun-etal-2023-chatgpt}. Specifically, the candidates in $\mathcal{C}_{p\%}(i)$ are divided into overlapping sliding windows, and each window is independently reranked by the LLM. The local ranking results are subsequently merged to produce a globally consistent ranking. Let $R(\cdot)$ denote the sliding-window reranking procedure. The initial top-$k$ candidates are obtained as
\begin{equation}
\mathcal{C}_{k}(i) = R\!\left(i,\mathcal{C}_{p\%}(i),M\right),
\end{equation}
where $M=\{M(c)\mid c\in\mathcal C_{p\%}(i)\}$ denotes the structured commit messages associated with the retrieved candidate commits. 
The resulting top-$k$ candidate commits constitute the initial traceability results. 

\subsection{Feedback-guided Refinement}\label{sec:refinement}

After obtaining the top-$k$ candidates from the initial traceability results, \toolname refines their \textit{Why} components using reasoning feedback and reranks the same candidate set. The initial message $M(c)$, including \textit{What} and \textit{Why}, is generated solely from the commit diff and can be reused across issue queries. While \textit{What} describes observable code changes and remains unchanged, the enhanced \textit{Why} is conditioned on the current issue $i$. Thus, the same commit may obtain different enhanced rationales for different issues, which are used specifically for second-round reranking rather than replacing the initial commit message.
Although the initial reranking narrows thousands of commits to a small set of promising candidates, distinguishing among them remains challenging. Due to the limited context in code changes, the initial \textit{Why} generated through one-shot inference may be inaccurate or overly general. We therefore introduce a \emph{Software Maintenance Knowledge Base} that captures transferable patterns from historical issue--commit links and provides additional evidence for feedback-guided \textit{Why} refinement.

\paragraph{Software Maintenance Knowledge Base Construction}

Rather than directly storing historical issue reports or commit messages, the proposed knowledge base organizes software maintenance experience into reusable maintenance patterns. Given a historical issue--commit link $(i,c)$, we first abstract the issue report into a generalized failure representation
\begin{equation}
A_I=\Phi_I(i),
\end{equation}
where $\Phi_I(\cdot)$ extracts transferable failure characteristics, including failure symptoms, execution context, affected entities, semantic clues, and possible failure mechanisms.

Similarly, instead of relying on implementation-specific code changes, we abstract the structured commit message $M(c)$ into a generalized repair representation
\begin{equation}
A_C=\Phi_C(M(c)),
\end{equation}
where $\Phi_C(\cdot)$ summarizes reusable repair knowledge, including repair operations, fix strategies, structural changes, semantic intent, and generalized root-cause categories.

The issue abstraction and commit abstraction are subsequently aligned to induce a transferable maintenance pattern
\begin{equation}
P=\Psi(A_I,A_C),
\end{equation}
where $\Psi(\cdot)$ denotes the pattern induction function. Each induced pattern describes the semantic relationship between generalized failure characteristics and corresponding repair strategies, together with an alignment rationale explaining why the repair resolves the underlying failure. The resulting knowledge base is represented as
\begin{equation}
\mathcal K=\{P_1,P_2,\cdots,P_n\},
\end{equation}
where each pattern represents reusable maintenance knowledge distilled from a historical issue--commit pair rather than repository-specific implementation details.

\paragraph{Pattern Retrieval and Reasoning Feedback}

Given a candidate issue--commit pair $(i,c)$ from the initial top-$k$ ranking $\mathcal{C}_{k}(i)$, the \textit{Feedback Agent} first constructs its abstract representation $(A_I,A_C)$ using the same abstraction functions. It then retrieves the most relevant maintenance patterns from $\mathcal K$
\begin{equation}
\mathcal P(i,c)=
Retrieve(A_I,A_C,\mathcal K),
\end{equation}
where $\mathcal P(i,c)$ denotes the retrieved maintenance patterns.

Instead of directly determining whether the candidate commit is correct, the retrieved patterns serve as historical maintenance evidence for the LLM judge, which generates reasoning feedback
\begin{equation}
F(i,c)=
J(A_I,A_C,\mathcal P(i,c)).
\end{equation}

\paragraph{Why Refinement and Second-round Reranking}

The reasoning feedback is subsequently returned to the \textit{Generation Agent} to refine the previously inferred \textit{Why} description $Y(c)$. 
Rather than treating the rationale inferred from commit diffs as a final explanation, \toolname regards $Y(c)$ as an initial hypothesis that can be validated and refined using traceability feedback. This design alleviates the risk of inaccurate rationale generation caused by limited context in commit diffs and enables the model to incorporate issue-related evidence when refining candidate commits. Accordingly, the enhanced structured commit message is represented as
\begin{equation}
M'(i,c)= \left\{W(c),Y'(i,c) \right\}, 
\end{equation}
\begin{equation}
Y'(i,c)= G_{\mathrm{refine}} \left(Y(c), F(i,c) \right),
\end{equation}
where the original \textit{What} remains unchanged and only the inferred \textit{Why} is refined based on the feedback associated with the current issue–commit pair. Accordingly, $M'(i,c)$ denotes an issue-conditioned enhanced commit message used for the current traceability query.

Finally, the enhanced structured commit messages are returned to the \textit{Traceability Agent} for a second-round reranking over the same top-$k$ candidate commits,
\begin{equation}
\mathcal C'_k(i)= R(i,\mathcal C_k(i),M'_i),
\end{equation}
where $M'_i=\{M'(i,c)\mid c\in\mathcal C_k(i)\}$ denotes the enhanced structured commit messages conditioned on issue $i$. Since the retrieval stage has achieved a favorable candidate recall–cost trade-off, the second-round reranking focuses solely on refining the relative ordering of these top-ranked candidates. The resulting ranking constitutes the final issue--commit traceability result, while the enhanced structured commit messages are simultaneously preserved as the final outputs of the proposed framework.

\section{Experimental Setup}\label{sec:experimental}
\subsection{Benchmark Construction}\label{sec:datasets}

\begin{table*}[h]
\centering
\small
\caption{Statistics of the curated benchmark.}
\label{tab:dataset}
\begin{tabular*}{\textwidth}{@{\extracolsep\fill}ccccccc}
\toprule
\multirow{2}{*}{\textbf{Project}} &
\textbf{Original} &
\textbf{Original} &
\textbf{\#Retained} &
\textbf{\#Retained} &
\textbf{Avg. Candidates} &
\textbf{Ground-truth} \\
& \textbf{Issues} & \textbf{Commits} & \textbf{Issues} & \textbf{Commits} & \textbf{per Issue} & \textbf{Links}\\
\midrule
Ambari   & 1000 & 24809  & 542 & 24295 & 3891.06 & 557 \\
Calcite  & 551  & 5889   & 447 & 5849 & 510.02 & 449 \\
Groovy   & 1000 & 20862  & 524 & 19813 & 975.40 & 544 \\
Ignite   & 1000 & 28869  & 566 & 22698 & 2203.39 & 590 \\
Isis     & 652  & 24945  & 327 & 22495 & 1183.77 & 359 \\
NetBeans & 159  & 10501  & 97 & 7040 & 874.34 & 101 \\
\bottomrule
\end{tabular*}
\begin{tablenotes}
\footnotesize
\item[a] \textbf{Note:} \textit{Original Issues} and \textit{Original Commits} are statistics from the original RDS benchmark~\cite{Huang2026Back}. \textit{Retained Issues} and \textit{Retained Commits} report the remaining issues and commits after benchmark curation. \textit{Avg. Candidates per Issue} denotes the average number of candidate commits per retained issue. \textit{Ground-truth Links} counts all positive issue--commit links after benchmark curation.
\end{tablenotes}
\end{table*}

We evaluate our approach on the issue--commit traceability benchmark released by Zhang et al.~\cite{zhang2023ealink}, which contains six Apache projects, namely \textit{Ambari}, \textit{Calcite}, \textit{Groovy}, \textit{Ignite}, \textit{Isis}, and \textit{NetBeans}. To construct a realistic candidate space, we follow the Realistic Distribution Setting (RDS) proposed by Huang et al.~\cite{Huang2026Back}, where candidate commits for each issue are collected based on the repository development activity after the issue creation time. Compared with randomly sampled settings, RDS preserves the naturally imbalanced candidate distribution encountered in real software repositories and therefore provides a more realistic evaluation setting for issue--commit traceability.
However, the dataset contains many ambiguous issue–commit links, making some problem–fix relationships difficult to interpret. Such ambiguity complicates determining whether generated commit messages capture the relevant problem and fix information through issue–commit traceability. We therefore further curate the benchmark by removing issues whose ground-truth links cannot be reliably interpreted, while preserving the original RDS candidate distribution for all retained issues.

The curation procedure consists of an automatic filtering stage followed by a manual inspection stage. During automatic filtering, we remove (1) issue--commit pairs whose commit diffs are unavailable, since the \textit{What} and \textit{Why} can only be generated from commit diffs, and (2) issues whose ground-truth commits are absent from the RDS candidate set, because such issues cannot be reliably evaluated under the traceability setting. The remaining issue--commit links are then manually reviewed to remove cases with unclear issue--commit correspondence. Specifically, we exclude two categories of samples:

\begin{itemize}
\item \textbf{Issues with ambiguous primary fixing commits.}
Some issues are linked to multiple commits without a clearly identifiable primary implementation change. Since no dominant problem--fix relationship can be consistently established, these issues are excluded.

\item \textbf{Issues with insufficient contextual information.}
Some issue reports contain only brief or incomplete descriptions, making it difficult to reliably interpret the semantic relationship between the reported problem and the corresponding code changes. Such issues are also excluded.
\end{itemize}

To minimize subjectivity, two authors independently reviewed all remaining issue--commit links and labeled each sample as either \textit{retain} or \textit{exclude}. Disagreements were resolved through discussion until consensus was reached.
The final benchmark contains 2,503 issues, 2,600 ground-truth links, and 102,190 unique candidate commits across the six projects. Detailed statistics are reported in Table~\ref{tab:dataset}.

\subsection{Baselines} \label{sec:baselines}

We construct four baseline pipelines by combining two CMG approaches with two issue--commit traceability approaches, all of which are state-of-the-art for their respective tasks. Each pipeline first generates a message for every candidate commit and then uses the generated message in place of the developer-written message required by the corresponding traceability approach. All other components of the traceability approach remain unchanged.

\textbf{(1) CMG approaches.}
\begin{itemize}
\item \textbf{ERICommiter}~\cite{xue2024automated} is a retrieval-based in-context learning approach. It employs a two-step filtering strategy to reduce the retrieval space, retrieves relevant historical diff--message pairs using lexical or semantic similarity, and incorporates the retrieved examples into the prompt to guide commit message generation.

\item \textbf{CoRaCMG}~\cite{xiong2026coracmg} is a contextual retrieval-augmented approach consisting of three stages: retrieval, augmentation, and generation. It uses a hybrid retriever to identify similar historical diff--message pairs, combines them with the target diff in a structured prompt, and generates the commit message from the augmented input.
\end{itemize}

\textbf{(2) Traceability approaches.}
\begin{itemize}
\item \textbf{LinkAnchor}~\cite{Akhavan2026LinkAnchor} is an autonomous LLM-based agent that formulates issue--commit traceability as a dynamic repository search process. It dynamically explores multiple repository artifacts and commit relationships to identify the resolving commit. 

\item \textbf{EasyLink}~\cite{Huang2026Back} is a training-free retrieve-and-rerank approach designed for the RDS setting. It first retrieves the top-10 candidate commits from a FAISS~\cite{douze2025faiss} vector database using all-MiniLM-L6-v2 embeddings, and then employs GPT-4o to rerank the retrieved candidates according to their semantic relevance to the issue report.
\end{itemize}

We combine each CMG approach with each traceability approach, resulting in four baselines: \textit{ERICommiter} + \textit{LinkAnchor}, \textit{ERICommiter} + \textit{EasyLink}, \textit{CoRaCMG} + \textit{LinkAnchor}, and \textit{CoRaCMG} + \textit{EasyLink}. For example, in \textit{ERICommiter} + \textit{LinkAnchor}, the developer-written messages originally used by LinkAnchor are replaced with messages generated by ERICommiter, while the remaining LinkAnchor workflow is preserved.

ERICommiter was originally evaluated with GPT-3.5 and Gemini, whereas CoRaCMG considered GPT-4o, GPT-4.1, DeepSeek-V3, DeepSeek-R1, and Qwen-2.5. Directly adopting these different backbones would make it difficult to distinguish the effects of the CMG frameworks from differences in model capability, especially given the rapid evolution of LLMs. Therefore, we use the same LLM backbone adopted in our study for ERICommiter, CoRaCMG, and \toolname, while retaining the retrieval, example construction, prompt augmentation, and generation procedures specific to each approach. This controlled setting helps reduce differences in backbone capability, making the comparison focus primarily on the CMG frameworks.

\subsection{Evaluation Metrics} \label{sec:evaluation metrics}

Following previous issue--commit traceability studies~\cite{Huang2026Back, zhang2023ealink, Akhavan2026LinkAnchor, radlinski2010comparing, sakai2008information}, we formulate issue--commit traceability as a ranking task and evaluate performance using two widely adopted metrics: Hit@$k$ and Mean Reciprocal Rank (MRR). 

\textbf{Hit@k.} Hit@$k$ measures whether at least one ground-truth commit is ranked within the top-$k$ candidates:
\begin{equation}
\mathrm{Hit@}k=
\frac{1}{|Q|}
\sum_{i=1}^{|Q|}
\mathbb{I}(Rank_i \le k),
\end{equation}
where $|Q|$ is the number of issues, and $Rank_i$ represents the rank position of the highest-ranked relevant commit for issue $i$. $\mathbb{I}(\cdot)$ is an indicator function that returns 1 if the condition is satisfied and 0 otherwise. We report Hit@1, Hit@5, and Hit@10 throughout the experiments.

\textbf{MRR.} MRR evaluates how highly the first relevant commit is ranked on average:
\begin{equation}
\mathrm{MRR}=
\frac{1}{|Q|}
\sum_{i=1}^{|Q|}
\frac{1}{Rank_i},
\end{equation}
where $Rank_i$ denotes the position of the first correctly linked commit in the ranked list for issue $i$. 


\subsection{Implementation Details}

All baselines and \toolname are implemented using Python 3.9.25 and PyTorch 2.4.1.
Experiments are conducted on a workstation equipped with 64 GB RAM, dual Intel Xeon Gold 6254 CPUs @ 3.10 GHz, and two NVIDIA GeForce RTX 4090 GPUs. The \textit{Generation Agent}, \textit{Traceability Agent}, and \textit{Feedback Agent} are all implemented using Qwen3.6-Plus through its official API with the temperature fixed at 0.1. Following EasyLink, dense retrieval is implemented using the Sentence-Transformers library with the all-MiniLM-L6-v2 encoder, where the top-$p\%$ retrieved candidates ($p=2$) are passed to the LLM ranker. The LLM ranker adopts the sliding-window strategy of RankGPT with a window size of 5 and a step size of 3. We set $k=10$ for the candidate set selected after the initial traceability reranking, which is subsequently used for feedback-guided refinement and second-round reranking. 
All inference parameters were selected on a validation subset sampled from the training split of the dataset released by Zhang et al.~\cite{zhang2023ealink} and fixed before evaluation. The benchmark used in our experiments was obtained by further curating the test split of that dataset, as described in Section~\ref{sec:datasets}. The sensitivity of the retrieval ratio and reranking parameters is analyzed in Section~\ref{sensitivity}. 
During feedback-guided refinement, the \textit{Feedback Agent} retrieves the top-3 most relevant maintenance patterns from the knowledge base for each issue--commit pair, and only a single round of feedback refinement is performed.
To improve efficiency, structured commit messages for all commits and the software maintenance knowledge base are generated offline and reused throughout all experiments. During inference, only knowledge retrieval, reasoning feedback, \textit{Why} refinement, and the second-round reranking are executed online. All prompts used by the three agents are included in the replication package to facilitate reproducibility.

\subsection{Research Questions}

In this study, we evaluate \toolname through four research questions:

\textbf{RQ1: How effective is \toolname compared with the baselines in serving issue--commit traceability?}

In this RQ, we evaluate the overall effectiveness of \toolname by comparing it with four combinations of representative CMG and traceability approaches: \textit{ERICommiter} + \textit{LinkAnchor}, \textit{ERICommiter} + \textit{EasyLink}, \textit{CoRaCMG} + \textit{LinkAnchor}, and \textit{CoRaCMG} + \textit{EasyLink}. Since LinkAnchor outputs only a single predicted resolving commit rather than a ranked candidate list, only Hit@1 is applicable to \textit{ERICommiter}+\textit{LinkAnchor} and \textit{CoRaCMG}+\textit{LinkAnchor}. Hit@1, Hit@5, Hit@10, and MRR are used for the remaining approaches.

\textbf{RQ2: How do different stages contribute to the effectiveness of \toolname?}

RQ2 investigates the contribution of each stage within \toolname by comparing its performance across three stage settings. Stage S1 denotes the ranking obtained by dense retrieval, which uses the issue report as the query and the structured messages produced by the \textit{Generation Agent} as the commit messages. Stage S2 denotes the ranking obtained by applying initial LLM-based reranking to the top-$p\%$ candidate commits retrieved at S1. Stage S3 denotes the final ranking obtained by refining the \textit{Why} components of the top-$k$ candidates from S2 using reasoning feedback and then performing second-round reranking. 

\textbf{RQ3: How do the generated \textit{What} and \textit{Why} contribute to issue--commit traceability?}

RQ3 aims to examine the contributions of the generated \textit{What} and \textit{Why}. We conduct an ablation study under the three stage settings introduced in RQ2: retrieval (Stage S1), initial reranking (Stage S2), and feedback-enhanced reranking (Stage S3). For each stage, we compare \textit{What only}, \textit{Why only}, and the complete \textit{What+Why} variant. In S3, \textit{Why only} and \textit{What+Why} apply feedback-guided \textit{Why} refinement before reranking. Since \textit{What only} contains no \textit{Why} component to refine, it skips the feedback stage and retains its Stage S2 results.

\textbf{RQ4: How does the practical usefulness of messages generated by \toolname compare with existing CMG approaches?}

To answer RQ4, we compare the practical usefulness of messages generated by \toolname, ERICommiter, and CoRaCMG through both LLM-as-a-Judge~\cite{zheng2023judging,kim2024prometheus,liu2023g} and human evaluation. We define four complementary dimensions: \textit{Change Specificity}, \textit{Cause Explanation}, \textit{Problem--Fix Alignment}, and \textit{Maintenance Usefulness}. Both evaluations follow the same protocol and rate each dimension on a five-point scale. A score of 1 indicates that the message provides little useful information or may be misleading, 3 indicates that it conveys useful information but omits important details, and 5 indicates that it provides comprehensive, accurate, and highly useful information.

\section{Results and Analysis}\label{sec:results}

\subsection{RQ1: How effective is \toolname compared with the baselines in serving issue--commit traceability?}\label{sec:rq1}

\begin{table*}[t]
\centering
\small
\caption{Overall comparison of traceability performance across six benchmark projects using different combinations of CMG and traceability approaches.}
\label{tab:overall_performance}
\begin{tabular*}{\textwidth}{@{\extracolsep\fill}cccccc}
\toprule
\textbf{Project} &
\textbf{Approach} &
\textbf{Hit@1} &
\textbf{Hit@5} &
\textbf{Hit@10} &
\textbf{MRR} \\
\midrule		
\multirow{5}{*}{Ambari}
& ERICommiter + EasyLink & 0.509 & 0.624 & 0.627 & \underline{0.569} \\
& ERICommiter + LinkAnchor & \underline{0.856} & - & - & - \\
& CoRaCMG + EasyLink & 0.177 & \underline{0.762} & \underline{0.935} & 0.407 \\
& CoRaCMG + LinkAnchor & 0.817 & - & - & - \\
& \textbf{\toolname} & \textbf{0.921}~(\(\uparrow\)\textbf{7.55\%}) & \textbf{0.935}~(\(\uparrow\)\textbf{22.76\%}) & \textbf{0.941}~(\(\uparrow\)\textbf{0.60\%}) & \textbf{0.930}~(\(\uparrow\)\textbf{63.24\%}) \\
\midrule	
\multirow{5}{*}{Calcite}
& ERICommiter + EasyLink & 0.716 & 0.812 & 0.814 & \underline{0.768} \\
& ERICommiter + LinkAnchor & \underline{0.919} & - & - & - \\
& CoRaCMG + EasyLink & 0.195 & \underline{0.817} & \underline{0.969} & 0.430 \\
& CoRaCMG + LinkAnchor & 0.832 & - & - & - \\
& \textbf{\toolname} & \textbf{0.944}~(\(\uparrow\)\textbf{2.68\%}) & \textbf{0.982}~(\(\uparrow\)\textbf{20.27\%}) & \textbf{0.984}~(\(\uparrow\)\textbf{1.61\%}) & \textbf{0.963}~(\(\uparrow\)\textbf{25.26\%}) \\
\midrule					
\multirow{5}{*}{Groovy}
& ERICommiter + EasyLink & 0.624 & 0.777 & 0.777 & \underline{0.697} \\
& ERICommiter + LinkAnchor & 0.712 & - & - & - \\
& CoRaCMG + EasyLink & 0.246 & \underline{0.809} & \underline{0.941} & 0.468 \\
& CoRaCMG + LinkAnchor & \underline{0.719} & - & - & - \\
& \textbf{\toolname} & \textbf{0.836}~(\(\uparrow\)\textbf{16.18\%}) & \textbf{0.914}~(\(\uparrow\)\textbf{12.96\%}) & \textbf{0.945}~(\(\uparrow\)\textbf{0.41\%}) & \textbf{0.871}~(\(\uparrow\)\textbf{25.04\%}) \\
\midrule		
\multirow{5}{*}{Ignite}
& ERICommiter + EasyLink & 0.564 & 0.684 & 0.686 & \underline{0.626} \\
& ERICommiter + LinkAnchor & \underline{0.758} & - & - & - \\
& CoRaCMG + EasyLink & 0.182 & \underline{0.758} & \underline{0.922} & 0.405 \\
& CoRaCMG + LinkAnchor & 0.744 & - & - & - \\
& \textbf{\toolname} & \textbf{0.882}~(\(\uparrow\)\textbf{16.31\%}) & \textbf{0.924}~(\(\uparrow\)\textbf{21.90\%}) & \textbf{0.924}~(\(\uparrow\)\textbf{0.18\%}) & \textbf{0.901}~(\(\uparrow\)\textbf{43.86\%}) \\
\midrule		
\multirow{5}{*}{Isis}
& ERICommiter + EasyLink & 0.505 & \underline{0.694} & 0.716 & \underline{0.590} \\
& ERICommiter + LinkAnchor & 0.615 & - & - & - \\
& CoRaCMG + EasyLink & 0.116 & 0.636 & \underline{0.869} & 0.329 \\
& CoRaCMG + LinkAnchor & \underline{0.654} & - & - & - \\
& \textbf{\toolname} & \textbf{0.829}~(\(\uparrow\)\textbf{26.63\%}) & \textbf{0.869}~(\(\uparrow\)\textbf{25.11\%}) & \textbf{0.872}~(\(\uparrow\)\textbf{0.36\%}) & \textbf{0.851}~(\(\uparrow\)\textbf{44.20\%}) \\
\midrule		
\multirow{5}{*}{NetBeans}
& ERICommiter + EasyLink & 0.619 & 0.753 & 0.753 & \underline{0.688} \\
& ERICommiter + LinkAnchor & 0.876 & - & - & - \\
& CoRaCMG + EasyLink & 0.289 & \underline{0.887} & \underline{0.949} & 0.519 \\
& CoRaCMG + LinkAnchor & \underline{0.907} & - & - & - \\
& \textbf{\toolname} & \textbf{0.949}~(\(\uparrow\)\textbf{4.55\%}) & \textbf{0.949}~(\(\uparrow\)\textbf{6.98\%}) & \textbf{0.949}~(\(\uparrow\)0.00\%) & \textbf{0.950}~(\(\uparrow\)\textbf{38.10\%}) \\
\midrule	
\multirow{5}{*}{Average}
& ERICommiter + EasyLink   & 0.589$^{\dagger L}$ & 0.724$^{\dagger L}$ & 0.729$^{\dagger L}$ & \underline{0.656}$^{\dagger L}$ \\
& ERICommiter + LinkAnchor & \underline{0.789}$^{\dagger L}$ & - & - & - \\
& CoRaCMG + EasyLink   & 0.201$^{\dagger L}$ & \underline{0.778}$^{\dagger L}$ & \underline{0.931}$^{S}$ & 0.426$^{\dagger L}$ \\
& CoRaCMG + LinkAnchor & 0.779$^{\dagger L}$ & - & - & - \\
& \textbf{\toolname} & \textbf{0.893}~(\(\uparrow\)\textbf{13.16\%}) & \textbf{0.929}~(\(\uparrow\)\textbf{19.37\%}) & \textbf{0.936}~(\(\uparrow\)\textbf{0.54\%}) & \textbf{0.911}~(\(\uparrow\)\textbf{38.75\%}) \\
\bottomrule
\end{tabular*}
\begin{tablenotes}
\footnotesize
\item[a] \textbf{Note:} 
$\dagger$ indicates a statistically significant improvement over the baseline according to the Wilcoxon signed-rank test ($p<0.05$)~\cite{wilcoxon1963critical}.
S, M, and L denote small, medium, and large effect sizes measured by Cliff's $\delta$~\cite{macbeth2011cliff}, respectively.
The best result for each metric is highlighted in bold, and the best baseline result is underlined. Percentages in parentheses show the relative improvements of \toolname over the best baseline for the corresponding metric.
\end{tablenotes}
\end{table*}

Table~\ref{tab:overall_performance} reports the performance of \toolname and the four baselines across the six projects. Overall, \toolname achieves the best average performance across all evaluation metrics, with average Hit@1, Hit@5, Hit@10, and MRR of 0.893, 0.929, 0.936, and 0.911, respectively. Compared with the best-performing baseline, the average improvements reach 13.16\%, 19.37\%, 0.54\%, and 38.75\%, respectively.
\toolname also achieves the highest Hit@1 on all six projects. According to the Wilcoxon signed-rank test and Cliff's $\delta$, the improvements over the best baseline are statistically significant ($p<0.05$) with large effect sizes for Hit@1, Hit@5, and MRR. The only exception is Hit@10 against \textit{CoRaCMG} + \textit{EasyLink}, for which the difference is not statistically significant and the effect size is small.

The four baselines exhibit distinct performance patterns depending on how the generated messages are used for traceability. 
First, when combined with EasyLink, ERICommiter and CoRaCMG perform differently across the evaluation metrics.
The baseline \textit{ERICommiter} + \textit{EasyLink} achieves higher Hit@1 and MRR, whereas \textit{CoRaCMG} + \textit{EasyLink} performs better on Hit@5 and Hit@10. One possible explanation lies in the different information
emphasized by their generated messages. In the baseline \textit{ERICommiter} + \textit{EasyLink}, the message generation component follows ERICommiter, which uses filtered historical diff--message pairs as in-context examples and emphasizes generating both \textit{What} and
\textit{Why} from the target diff. Although its development intent is inferred solely from the diff, this information may provide additional semantic cues for placing the fixing commit at the top of the ranking. In contrast, the message generation component of the baseline \textit{CoRaCMG} + \textit{EasyLink} follows CoRaCMG, which uses a hybrid retriever to obtain similar historical diff--message pairs and guides the LLM to capture project-specific terminology, writing styles, and implementation details. Its higher Hit@5 and Hit@10 suggest that these detailed change descriptions are effective in placing relevant commits within the top-ranked candidate set. However, when several top-ranked commits modify similar files, locations, or functionality,
implementation-focused descriptions may provide limited evidence for distinguishing their underlying development intent. This may explain why \textit{CoRaCMG} + \textit{EasyLink} achieves relatively lower Hit@1 and
MRR. 

Second, when combined with LinkAnchor, the two baselines
\textit{ERICommiter} + \textit{LinkAnchor} and \textit{CoRaCMG} + \textit{LinkAnchor} achieve similar Hit@1 results of 0.789 and 0.779, respectively. Their difference is only 0.010, whereas the difference between
\textit{ERICommiter} + \textit{EasyLink} and \textit{CoRaCMG} + \textit{EasyLink} reaches 0.388. In \textit{ERICommiter} + \textit{EasyLink} and \textit{CoRaCMG} + \textit{EasyLink}, candidate ranking mainly depends on the direct semantic relationship between the issue report and the generated message. The
different information emphasized by the two message generation components may therefore have a greater influence on the ranking results. In \textit{ERICommiter} + \textit{LinkAnchor} and \textit{CoRaCMG} + \textit{LinkAnchor}, generated messages participate in an agent-based search process that uses their change and intent information to identify relevant commits and examine their relationships with the reported issue. The agent can then contextualize and verify this information by selectively examining issue discussions, commit histories, and code changes. This multi-step reasoning process may enable LinkAnchor to make effective use of the
information emphasized by either CMG approach, thereby reducing the performance difference between the two combinations.

Compared with the four baselines, \toolname more effectively serves issue--commit traceability by integrating message generation with traceability reasoning. In the baseline combinations, commit messages are generated before traceability and remain unchanged during candidate ranking. Consequently, development intent inferred solely from a diff may be general
or ambiguous, while messages emphasizing implementation details may provide insufficient evidence for distinguishing commits with similar changes. \toolname first uses the generated \textit{What} and \textit{Why} to serve candidate retrieval and reranking, and then uses the traceability results to guide the refinement of \textit{Why} for the top-ranked candidates. This
interaction preserves the implementation information in \textit{What} while making the development intent more specific to the current issue. The refined problem--fix information enables \toolname to better distinguish the actual fixing commit, leading to consistent improvements in Hit@1 and MRR.

\begin{rqsummary}
\textbf{RQ1 Summary.}
\toolname achieves the best overall performance by enabling generated messages to serve issue--commit traceability and using traceability feedback for issue-conditioned refinement, demonstrating the effectiveness of the proposed collaborative multi-agent framework for this dynamic process in AI-assisted workflows.
\end{rqsummary}

\subsection{RQ2: How do different stages contribute to the effectiveness of \toolname?}

Table~\ref{tab:stage_wise_comparison} reports the average results and incremental gains. From S1 to S2, the average Hit@1, Hit@5, Hit@10, and MRR increase from 0.488, 0.751, 0.838, and 0.609 to 0.697, 0.917, 0.936, and 0.800, corresponding to gains of 0.209, 0.166, 0.098, and 0.191, respectively. After feedback-guided refinement at S3, Hit@1 and MRR further increase to 0.893 and 0.911, with gains of 0.196 and 0.111 over S2. 

\begin{table}[t]
\centering
\small
\caption{Average traceability performance and incremental gains of \toolname across the three stages.}
\label{tab:stage_wise_comparison}
\begin{tabular*}{\linewidth}{@{\extracolsep{\fill}}cccc|cc@{}}
\toprule
\multirow{2}{*}{\textbf{Metric}} &
\multicolumn{3}{c|}{\textbf{\toolname}} &
\multicolumn{2}{c}{\textbf{Stage Gain}} \\
\cmidrule(lr){2-4}
\cmidrule(lr){5-6}
 &
\textbf{S1} &
\textbf{S2} &
\textbf{S3} &
\textbf{$G_{S2-S1}$} &
\textbf{$G_{S3-S2}$} \\
\midrule					
Hit@1  & 0.488 & 0.697 & 0.893 & + 0.209 & + 0.196 \\
Hit@5  & 0.751 & 0.917 & 0.929 & + 0.166 & + 0.012 \\
Hit@10 & 0.838 & 0.936 & 0.936 & + 0.098 & + 0.000 \\
MRR    & 0.609 & 0.800 & 0.911 & + 0.191 & + 0.111 \\
\bottomrule
\end{tabular*}
\begin{tablenotes}
\footnotesize
\item \textbf{Note:} $G_{S2-S1}$ and $G_{S3-S2}$ represent the incremental performance gains achieved by LLM reranking and feedback-guided refinement within \toolname, respectively.
\end{tablenotes}
\end{table}

The performance differences among the three stages reflect how the generated messages are utilized. At S1, dense retrieval is implemented using a lightweight embedding model with limited ability to capture complex problem--fix semantics. Its ranking is therefore more dependent on surface-level textual overlap. However, the structured messages are generated by the \textit{Generation Agent} through semantic reasoning over commit diffs rather than by maximizing keyword overlap with issue reports, making their semantic information difficult to fully exploit through dense retrieval alone. At S2, the LLM-based reranker jointly reasons over the issue report and the generated \textit{What} and \textit{Why}. The two components provide richer evidence about implementation changes and development intent, allowing the LLM to better distinguish relevant commits from semantically similar candidates and substantially improve the ranking. At S3, the \textit{Why} of each top-$k$ candidate is further refined according to the current issue and reasoning feedback derived from historical maintenance patterns. This refinement makes the inferred development intent more specific to the reported problem, increasing Hit@1 from 0.697 to 0.893 and MRR from 0.800 to 0.911. It therefore helps to distinguish the actual fixing commit from multiple closely related candidates already identified in the preceding stages.

Overall, S1 establishes the initial ranking using the generated structured messages, S2 enables their semantic information to be more effectively interpreted through LLM-based reasoning, and S3 further improves their problem--fix alignment through issue-conditioned \textit{Why} refinement. These results demonstrate that message generation, traceability reasoning, and feedback-guided refinement make complementary contributions to the effectiveness of \toolname.

\begin{rqsummary}
\textbf{RQ2 Summary.}
LLM-based reranking effectively utilizes the semantic evidence contained in the generated \textit{What} and \textit{Why}, while feedback-guided refinement further improves top-ranked performance by making the inferred development intent more specific to the current issue.
\end{rqsummary}

\subsection{RQ3: How do the generated \textit{What} and \textit{Why} contribute to issue--commit traceability?}

\begin{table*}[t]
\centering
\small
\caption{Ablation study of the generated \textit{What} and \textit{Why} under different inference settings.}
\label{tab:ablation}
\begin{tabular*}{\textwidth}{@{\extracolsep\fill}cccl|cl|cl|cl}
\toprule
\textbf{Stage} &
\textbf{Variant} &
\multicolumn{2}{c|}{\textbf{Hit@1}} &
\multicolumn{2}{c|}{\textbf{Hit@5}} &
\multicolumn{2}{c|}{\textbf{Hit@10}} &
\multicolumn{2}{c}{\textbf{MRR}} \\
\midrule
\multirow{3}{*}{S1}
& What only
& 0.409 & $\downarrow$ 16.2\% & 0.686 & $\downarrow$ 8.6\% & 0.780 & $\downarrow$ 6.9\% & 0.536 & $\downarrow$ 11.9\% \\
& Why only
& 0.450 & $\downarrow$ 7.7\% & 0.664 & $\downarrow$ 11.6\% & 0.726 & $\downarrow$ 13.3\% & 0.549 & $\downarrow$ 9.8\% \\
& \textbf{What+Why}
& \textbf{0.488} & -- & \textbf{0.751} & -- & \textbf{0.838} & -- & \textbf{0.609} & -- \\
\midrule
\multirow{3}{*}{S2}
& What only
& 0.511 & $\downarrow$ 26.8\% & 0.800 & $\downarrow$ 12.8\% & 0.847 & $\downarrow$ 9.5\% & 0.640 & $\downarrow$ 19.9\% \\
& Why only
& 0.544 & $\downarrow$ 22.1\% & 0.741 & $\downarrow$ 19.2\% & 0.780 & $\downarrow$ 16.6\% & 0.634 & $\downarrow$ 20.6\% \\
& \textbf{What+Why}
& \textbf{0.697} & -- & \textbf{0.917} & -- & \textbf{0.936} & -- & \textbf{0.800} & -- \\
\midrule
\multirow{3}{*}{S3}
& What only
& 0.511 & $\downarrow$ 42.8\% & 0.800 & $\downarrow$ 13.9\% & 0.847 & $\downarrow$ 9.5\% & 0.640 & $\downarrow$ 29.6\% \\
& Why only
& 0.684 & $\downarrow$ 23.4\% & 0.761 & $\downarrow$ 18.1\% & 0.780 & $\downarrow$ 16.6\% & 0.718 & $\downarrow$ 21.2\% \\
& \textbf{What+Why}
& \textbf{0.893} & -- & \textbf{0.929} & -- & \textbf{0.936} & -- & \textbf{0.911} & -- \\
\bottomrule
\end{tabular*}
\end{table*}

Table~\ref{tab:ablation} reports the ablation results for the generated \textit{What} and \textit{Why}.
Overall, removing either message component consistently degrades traceability performance across all three stage settings, while combining both components (\textit{What+Why}) consistently achieves the best results. 
Compared with \textit{What+Why}, removing \textit{Why} results in relative Hit@1 decreases of 16.2\%, 26.8\%, and 42.8\% at S1--S3, respectively. Removing \textit{What} produces corresponding decreases of 7.7\%, 22.1\%, and 23.4\%. Similar performance decreases are observed for Hit@5, Hit@10, and MRR, demonstrating that implementation changes and development intent provide complementary evidence for issue--commit traceability.

Comparing the two ablated variants \textit{What only} and \textit{Why only} further reveals their different contributions. At S1, \textit{Why only} exceeds \textit{What only} in Hit@1 by 0.041, whereas \textit{What only} achieves higher Hit@5 and Hit@10 by 0.022 and 0.054, respectively. At S2, \textit{Why only} remains higher in Hit@1 by 0.033, while \textit{What only} performs slightly better in Hit@5, Hit@10, and MRR by 0.059, 0.067, and 0.006. At S3, the Hit@1 difference in favor of \textit{Why only} increases to 0.173, whereas \textit{What only} remains higher in Hit@5 and Hit@10 by 0.039 and 0.067. These results indicate that \textit{Why} is particularly important for identifying the actual fixing commit at the top position, while \textit{What} provides broader coverage among the top-ranked candidates.

These differences arise from the distinct information captured by the two components. \textit{What} summarizes implementation changes and preserves project-specific entities such as files, classes, and methods, providing lexical and structural evidence for retrieving potentially relevant commits. In contrast, \textit{Why} describes the underlying development intent and repair rationale, providing problem--fix information that helps distinguish the actual fixing commit from other candidates with similar code changes. LLM-based reranking can make greater use of this rationale-level evidence than dense retrieval, while feedback-guided refinement further makes the generated \textit{Why} specific to the current issue. This explains why the contribution of \textit{Why} to Hit@1 becomes more pronounced at S2 and S3.

\begin{rqsummary}
\textbf{RQ3 Summary.}
\textit{What} and \textit{Why} provide complementary evidence for issue--commit traceability. Their combination consistently achieves the best performance, with \textit{Why} contributing more to identifying the actual fixing commit at the top position and \textit{What} supporting broader top-ranked candidate coverage.
\end{rqsummary}

\subsection{RQ4: How does the practical usefulness of messages generated by \toolname compare with existing CMG approaches?}

\begin{figure*}[!t]
  \centering
 \includegraphics[width=\linewidth]{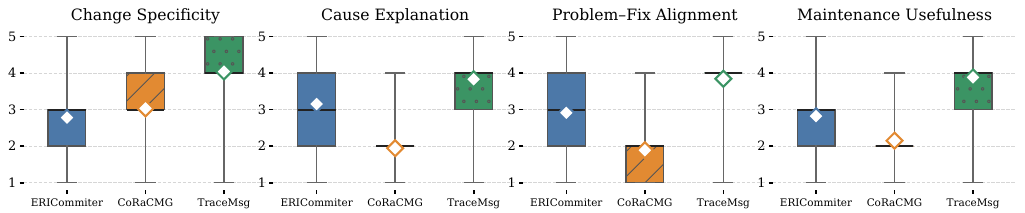}
  \caption{Distribution of LLM-as-a-Judge scores for commit messages generated by different CMG approaches across the six benchmark projects.}
  \label{fig:rq4}
\end{figure*}

\begin{table*}[t]
\centering
\caption{Average LLM-as-a-Judge and human evaluation scores of commit messages generated by different CMG approaches.}
\label{tab:CMG}
\small
\setlength{\tabcolsep}{9pt}
\begin{tabular*}{\textwidth}{@{\extracolsep\fill}cccc|ccc}
\toprule
\multirow{2}{*}{\textbf{Metric}}
& \multicolumn{3}{c|}{\textbf{LLM-based Evaluation}}
& \multicolumn{3}{c}{\textbf{Human Evaluation}} \\
\cmidrule(lr){2-4}
\cmidrule(lr){5-7}
& \textbf{ERICommiter}
& \textbf{CoRaCMG}
& \textbf{\toolname}
& \textbf{ERICommiter}
& \textbf{CoRaCMG}
& \textbf{\toolname} \\
\midrule
Change Specificity
& 2.76 & 3.02 & \textbf{4.03}
& 2.42 & 3.06 & \textbf{4.13} \\
Cause Explanation
& 3.13 & 1.93 & \textbf{3.85}
& 3.09 & 1.85 & \textbf{3.71} \\
Problem--Fix Alignment
& 2.87 & 1.88 & \textbf{3.85}
& 3.08 & 2.14 & \textbf{3.79} \\
Maintenance Usefulness
& 2.80 & 2.15 & \textbf{3.88}
& 3.01 & 2.37 & \textbf{3.83} \\
\midrule
\textbf{Average}
& 2.89 & 2.25 & \textbf{3.90} 
& 2.90 & 2.36 & \textbf{3.87} \\
\bottomrule
\end{tabular*}
\end{table*}

We employ GLM-5.1 as an LLM-as-a-Judge evaluator to assess the commit messages of all issue-linked commits included in the benchmark. 
For each issue–commit pair, the evaluator is provided with the issue report, commit diff, and commit message, while the identity of the CMG approach is anonymized.
To further validate the reliability of the LLM-as-a-Judge assessment, we conduct a source-anonymized human evaluation on a representative subset of the benchmark. We randomly sample 20 issue--commit pairs from each of the six projects, resulting in 120 sampled cases in total. The evaluation involves six participants, including two Ph.D. students, two master's students, and two professional software developers, with 3--8 years of software development experience. Each sampled commit message is independently evaluated by three participants, and the final human score is calculated as the average of the three ratings. 

Fig.~\ref{fig:rq4} presents the score distributions from the full LLM-based evaluation, while Table~\ref{tab:CMG} reports the average results of both the LLM-based and human evaluations. The human evaluation further supports the reliability of the LLM-based results. For each message and evaluation dimension, the three human ratings are averaged and rounded to the nearest integer before being compared with the corresponding LLM rating. The resulting Cohen's kappa~\cite{chmura2002kappa} score of 0.83 indicates strong agreement between the two evaluations.

As shown in Table~\ref{tab:CMG}, \toolname achieves the highest average scores under both evaluation methods. Its overall average reaches 3.90 in the LLM-based evaluation, compared with 2.89 for ERICommiter and 2.25 for CoRaCMG. The human evaluation produces consistent results, with average scores of 3.87, 2.90, and 2.36 for \toolname, ERICommiter, and CoRaCMG, respectively. \toolname also ranks first in all four dimensions. Fig.~\ref{fig:rq4} further shows that its scores are concentrated at the upper end of the scale. In particular, its ratings are primarily distributed between 4 and 5 for \textit{Change Specificity}, and around 4 for \textit{Cause Explanation} and \textit{Problem--Fix Alignment}. Its \textit{Maintenance Usefulness} scores are also mainly distributed between 3 and 4.

The differences among the three approaches can be explained by how they generate commit messages. CoRaCMG prompts the LLM with a retrieved diff--message exemplar and asks it to generate a concise message for the target diff. This design primarily guides the model to describe the code modifications and follow the terminology and writing style of the retrieved example. Consequently, CoRaCMG performs relatively well in \textit{Change Specificity}, achieving 3.02 and 3.06 under the LLM-based and human evaluations, but receives substantially lower scores in \textit{Cause Explanation} and \textit{Problem--Fix Alignment}. This result illustrates that accurately describing a diff does not necessarily ensure that the generated message conveys the information required by other software engineering activities.
ERICommiter explicitly guides the LLM to generate both \textit{What} and \textit{Why} information from the commit diff. It therefore provides stronger cause-related information than CoRaCMG, achieving LLM-based scores of 3.13 for \textit{Cause Explanation} and 2.87 for \textit{Problem--Fix Alignment}. However, because its \textit{Why} is inferred solely from the diff, the resulting rationale may remain general or fail to identify the specific problem that motivates the change. By contrast, \toolname refines the generated \textit{Why} through one round of traceability-guided feedback conditioned on the current issue. This process makes the development intent more specific and strengthens the problem--fix relationship expressed by the message, leading to consistently higher scores across all four dimensions.

\begin{figure*}[!t]
  \centering
 \includegraphics[width=\linewidth]{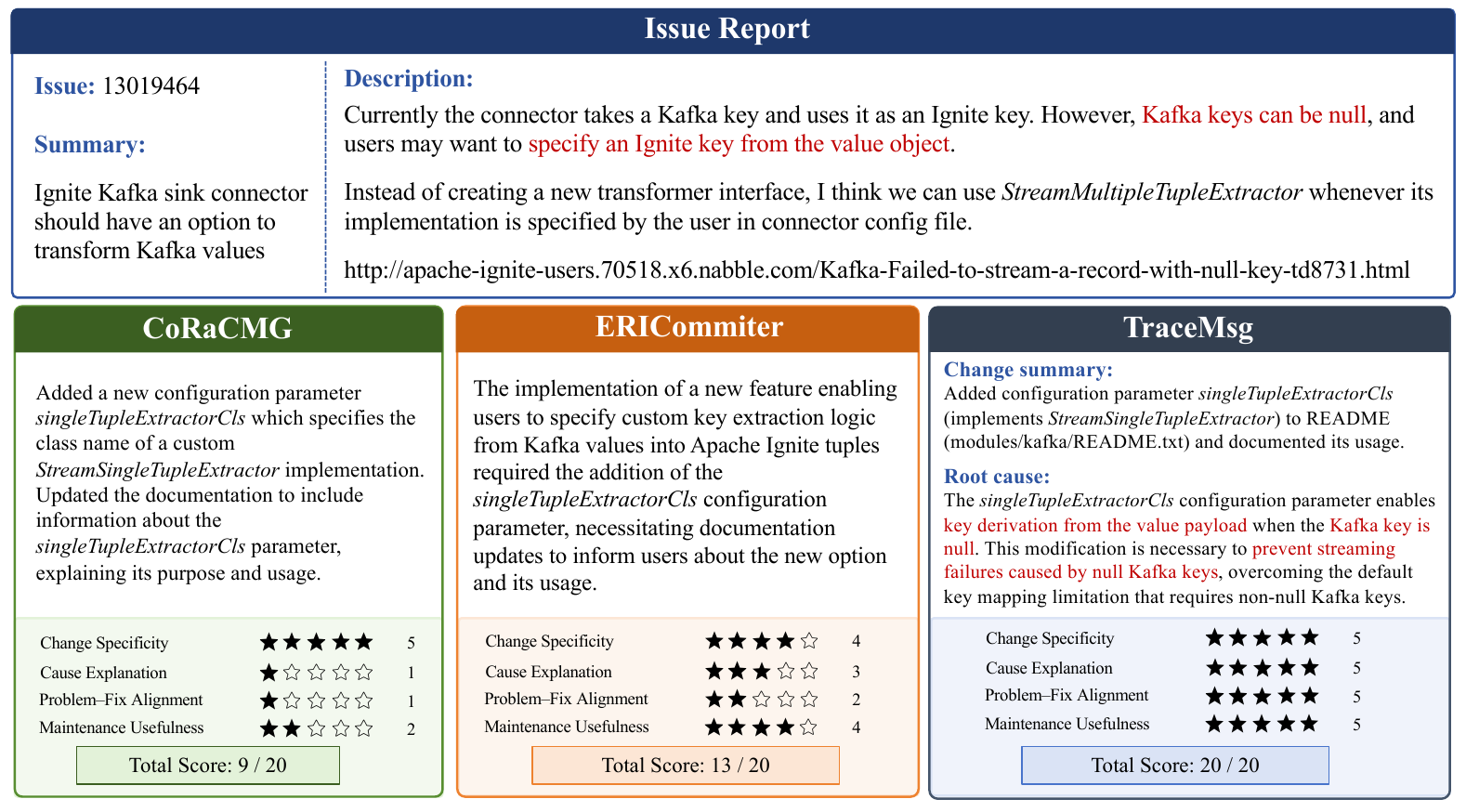}
 \caption{An illustrative case from the \textit{Apache Ignite} project comparing the development information conveyed by messages generated by different CMG approaches.}
  \label{fig:rq_case}
\end{figure*}

Fig.~\ref{fig:rq_case} presents an illustrative case from the \textit{Ignite} project to show how messages generated by different CMG approaches convey development information. The issue reports that Kafka keys can be null and requests an option to derive an Ignite key from the value object. CoRaCMG accurately describes the addition and documentation of the \texttt{singleTupleExtractorCls} parameter, but does not explain the reported problem or how the new parameter resolves it. It therefore receives 5 for \textit{Change Specificity}, but only 1, 1, and 2 for the other three dimensions. ERICommiter further infers that the parameter enables custom key extraction from Kafka values, but it does not explicitly identify null Kafka keys as the cause of streaming failures, resulting in a total score of 13 out of 20. In contrast, \toolname connects the new parameter with key derivation from the value payload when the Kafka key is null and explains that this change prevents the resulting streaming failures. It consequently receives the highest score in all four dimensions, with a total score of 20 out of 20.

\begin{rqsummary}
\textbf{RQ4 Summary.}
\toolname improves the practical usefulness of generated messages by providing clearer causes and problem--fix relationships beyond code-change descriptions.
\end{rqsummary}

\section{Discussion}\label{sec:discussion}

\subsection{Sensitivity Analysis of Retrieval and Reranking Parameters}\label{sensitivity}

Although \toolname is a training-free framework, two inference parameters need to be determined before evaluation: the retrieval ratio used for candidate selection and the sliding-window configuration used during LLM reranking. To avoid information leakage, we do not use the test split for parameter selection. Instead, following the official benchmark partition, the training split is used solely as a source of historical issue--commit pairs for knowledge base construction and parameter validation, rather than model training. Specifically, we randomly sample 50 issue--commit pairs from each project (300 pairs in total) using a fixed random seed, and all parameter settings are selected exclusively on this validation subset before being fixed throughout the remaining experiments.

\textbf{Retrieval ratio.}
Fig.~\ref{fig:discussion_analysis}(a) shows the retrieval recall achieved under different candidate ratios. As expected, retaining more candidates consistently improves retrieval recall, increasing from 78.4\% at top-0.5\% to 93.2\% at top-2\%. Beyond this point, however, the improvement becomes marginal. Expanding the candidate pool from top-2\% to top-5\% increases the reranking workload by approximately 2.5 times, while improving retrieval recall by only 3.9\%. Since the LLM reranking stage dominates the overall inference cost, the additional computational overhead outweighs the limited recall gain. Therefore, we adopt a retrieval ratio of top-2\%, which provides a favorable balance between retrieval coverage and reranking efficiency.

\textbf{Sliding-window configuration.}
Fig.~\ref{fig:discussion_analysis}(b) evaluates different combinations of window sizes and step sizes in the RankGPT-style reranking strategy. Following RankGPT, the step size is fixed to approximately half of the corresponding window size. Overall, the performance remains relatively stable across different parameter settings, indicating that the reranking stage is not overly sensitive to the exact window configuration. Among all evaluated settings, $(w,s)=(5,3)$ consistently achieves the best overall performance on both Hit@1 and MRR. Smaller windows limit the number of candidates that can be compared simultaneously, whereas larger windows introduce substantially longer contexts with only marginal ranking improvements. Consequently, $(w,s)=(5,3)$ is adopted throughout this work.

The proposed framework is relatively robust to different retrieval and reranking parameters. The selected configuration (top-2\% retrieval and $(w,s)=(5,3)$) achieves a favorable trade-off between effectiveness and inference efficiency while remaining stable across nearby parameter settings.

\subsection{Robustness to Different Backbone LLMs}\label{llm_robustness}

Since \toolname relies on LLMs to generate development knowledge, an important question is whether its effectiveness depends heavily on the choice of the backbone model. To answer this question, we evaluate four representative frontier LLMs, including Qwen3.6-Plus, GPT-5.2, GLM-5.1, and DeepSeek-V3.
To isolate the influence of the generated \textit{What} and \textit{Why}, we replace only the \textit{Generation Agent} while keeping the subsequent traceability pipeline unchanged. Specifically, the same retrieval, reranking, and feedback-guided inference processes are adopted for all backbone models. Thus, differences in the results mainly reflect changes in the generated \textit{What} and \textit{Why} descriptions. To control API costs while preserving the project distribution of the benchmark, we randomly sample 10\% of the test issues from each project, resulting in 251 issues. Hit@1 and MRR are adopted as the evaluation metrics.

\begin{figure*}[!t]
    \centering
    \subfloat[
        \label{fig:discussion_retrieval}
    ]{\includegraphics[width=0.32\textwidth]
        {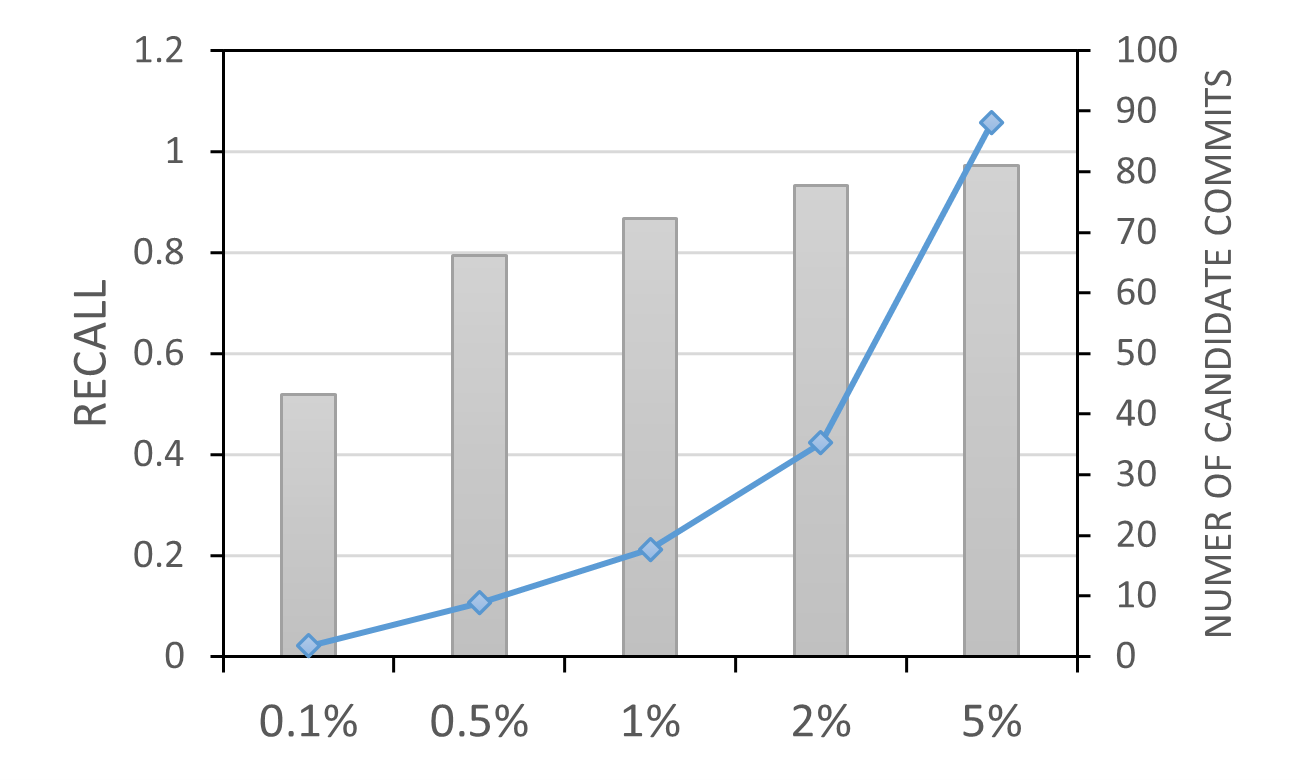}}
    \hfill
    \subfloat[
        \label{fig:discussion_window}
    ]{\includegraphics[width=0.32\textwidth]
        {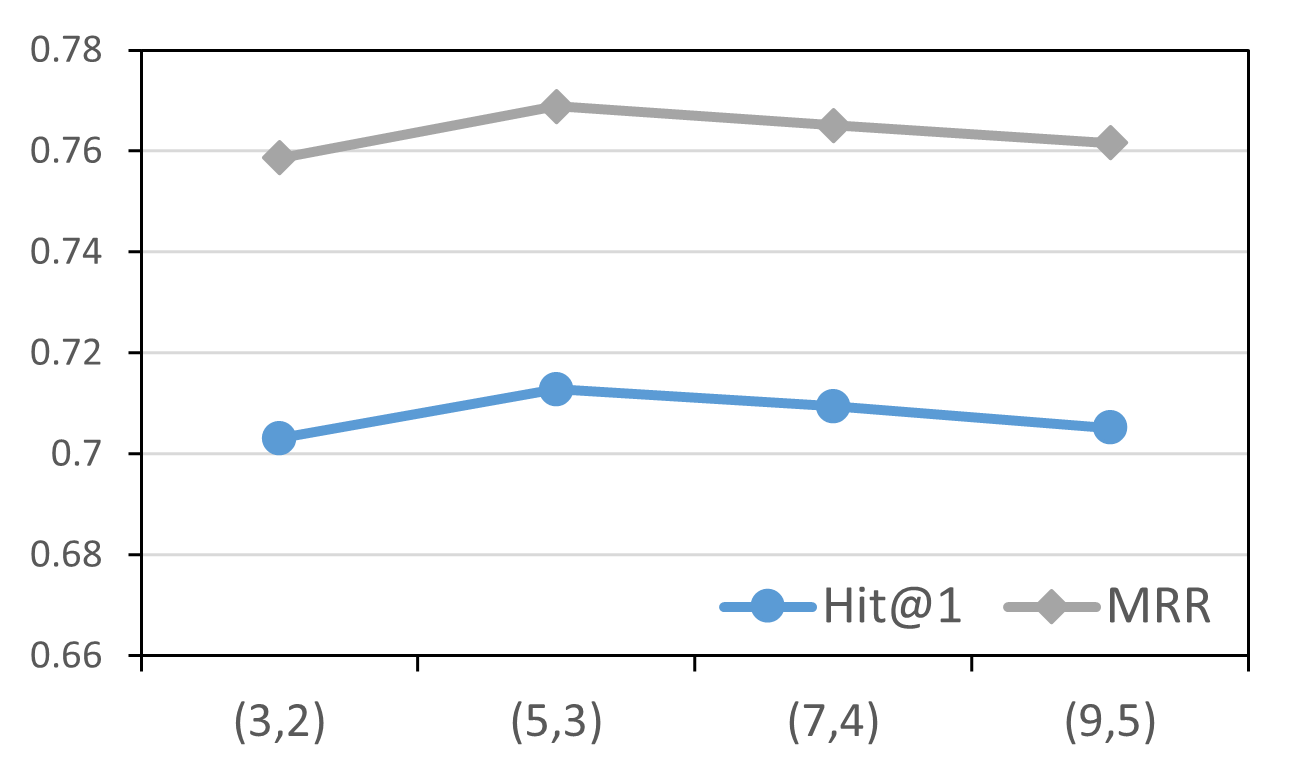}}
    \hfill
    \subfloat[
        \label{fig:discussion_llm}
    ]{\includegraphics[width=0.32\textwidth]
        {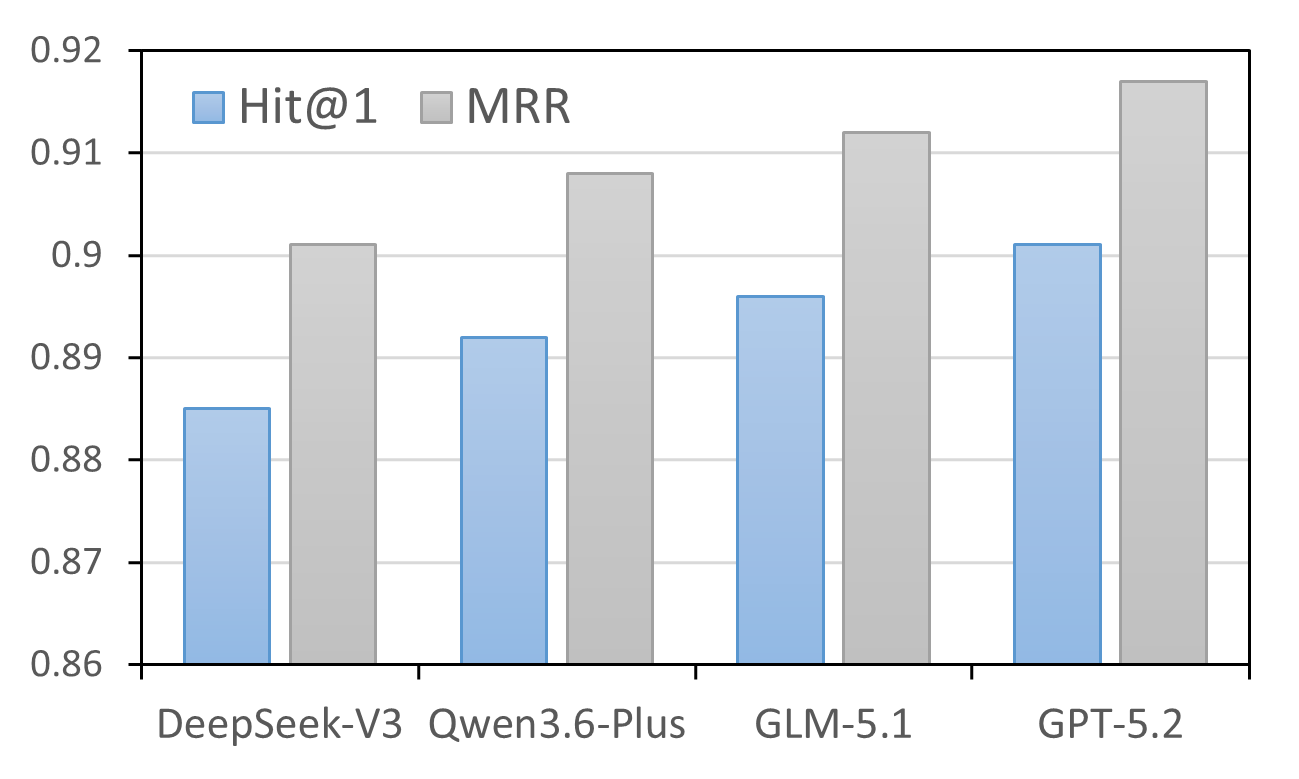}}
    \caption{Parameter sensitivity and backbone LLM analysis of \toolname.
    (a) Retrieval recall and the average number of candidate commits under different retrieval ratios.
    (b) Hit@1 and MRR under different sliding-window configurations.
    (c) Traceability performance with different backbone LLMs.}
    \label{fig:discussion_analysis}
\end{figure*}

Fig.~\ref{fig:discussion_analysis}(c) summarizes the results. GPT-5.2 achieves the best overall performance, obtaining a Hit@1 of 90.1\% and an MRR of 91.7\%, followed by GLM-5.1 (89.6\%/91.2\%), Qwen3.6-Plus (89.2\%/90.8\%), and DeepSeek-V3 (88.5\%/90.1\%). Although stronger backbone models consistently produce slightly better traceability performance, the performance differences remain relatively small. The maximum gap between the best and worst backbone models is only 1.6 percentage points on Hit@1 and 1.6 percentage points on MRR.
These results suggest that \toolname is not highly sensitive to the evaluated backbone choices.
More capable backbone models improve the quality of the generated \textit{What} and \textit{Why}, leading to incremental gains in issue--commit traceability, while the overall framework remains robust across different backbone choices. Considering that GPT-5.2 and GLM-5.1 incur substantially higher API costs than Qwen3.6-Plus according to their official pricing, yet provide only marginal performance improvements, we adopt Qwen3.6-Plus as the default backbone throughout this work because it offers the most favorable balance between effectiveness and deployment cost.

\subsection{Efficiency and Cost Analysis}\label{efficiency}

\begin{table*}[t]
\centering
\caption{Average runtime and API cost of each component.}
\label{tab:cost}
\small
\begin{tabular*}{\textwidth}{@{\extracolsep\fill}cccccccc}
\toprule
\multirow{2}{*}{\textbf{Component}} &
\multirow{2}{*}{\textbf{Execution}} &
\multirow{2}{*}{\textbf{Unit}} &
\multicolumn{1}{c}{\textbf{Avg. Input}} &
\multicolumn{1}{c}{\textbf{Avg. Output}} &
\multicolumn{1}{c}{\textbf{API}} &
\multicolumn{1}{c}{\textbf{Avg.}} &
\multicolumn{1}{c}{\textbf{Avg.}} \\
& & & \textbf{Tokens} & \textbf{Tokens} & \textbf{Calls} & \textbf{Time (s)} & \textbf{Cost (\$)} \\
\midrule
Generation Agent & Offline & Commit
& 5,655 & 1,985 & 1 & 5.1 & 0.009 \\
Traceability Agent & Online & Issue
& 121,979 & 440 & 12 & 18.7 & 0.062 \\
Feedback Agent & Online & Issue
& 47,050 & 8,250 & 13 & 24.5 & 0.048 \\
\bottomrule
\end{tabular*}
\end{table*}

Although \toolname introduces an additional feedback-guided inference step compared with conventional retrieve-and-rerank frameworks, its computational overhead remains practical because initial commit message generation is performed offline, and the resulting messages are reused across subsequent issue queries.
Table~\ref{tab:cost} summarizes the average runtime, token consumption, API calls, and monetary cost of each component.

\textbf{Offline preprocessing.}
The semantic descriptions (\textit{What} and \textit{Why}) are generated once for each commit during repository preprocessing. On average, each commit contains 5,655 input tokens and produces 1,985 output tokens through a single LLM call, requiring approximately 5.1 seconds and \$0.009 per commit. Since these descriptions are generated only once and reused for all future issue--commit traceability tasks, this cost is amortized over repeated online queries and therefore does not affect online inference efficiency.

\textbf{Online inference.}
For each issue, dense retrieval is first performed using the lightweight all-MiniLM-L6-v2 encoder without invoking any LLM. Only the top-2\% retrieved candidates (35.3 commits on average) are forwarded to the reranking stage. Using the adopted sliding-window strategy $(w,s)=(5,3)$, reranking requires approximately 12 API calls per issue. The subsequent feedback-guided inference further refines the generated \textit{Why} for only the top-10 reranked candidates before performing a second-round reranking, resulting in another 13 API calls. Consequently, the additional reasoning introduced by \toolname is restricted to a very small subset of highly relevant candidates instead of the entire retrieval space.
Overall, the complete online inference requires approximately 25 API calls, processes about $1.78\times10^5$ input and output tokens, costs approximately \$0.11 per issue under the official Qwen3.6-Plus pricing, and takes 43.2 seconds on average. Among all components, LLM reranking accounts for the largest proportion of the inference time, whereas the additional feedback-guided refinement introduces only a moderate increase in computational cost while providing substantial improvements in traceability performance.

\section{Threats to Validity}\label{sec:threats}

Our study is subject to two main threats to validity. The first threat concerns the reliability and representativeness of the benchmark annotations. Although our benchmark is constructed based on the issue--commit links provided by the EasyLink~\cite{Huang2026Back} dataset, ambiguous links may still exist in real-world repositories, particularly for issues involving multiple related commits or repository-wide refactorings. To mitigate this threat, we performed an additional manual curation process to remove ambiguous samples before evaluation. Two authors independently inspected the issue descriptions, commit metadata, and code changes, and resolved disagreements through discussion. As a result, the curated benchmark provides a more reliable evaluation setting and reduces the impact of annotation ambiguity on the reported results. 
Excluding issues with ambiguous links or insufficient contextual information may also bias the benchmark toward cases with clearer problem--fix relationships. TraceMsg and all baselines were evaluated on the same curated issues, so the filtering did not give TraceMsg a different set of evaluation samples and supports a fair comparison.

The second threat concerns the generalizability of our experimental findings. Our evaluation is conducted on six Java projects collected from Apache repositories. Although these projects cover diverse application domains and development histories, the results may not directly generalize to other programming languages, issue tracking systems, or industrial software projects. Nevertheless, the proposed framework is independent of language-specific features and can be applied to other repositories. In addition, we evaluate the framework using multiple backbone LLMs and observe consistently strong performance across different models. Detailed results are provided in Section~\ref{llm_robustness}.

\section{Related Work}\label{sec:related_work}
\subsection{Commit Message Generation}

Commit message generation (CMG) aims to automatically generate natural-language descriptions of code changes to facilitate software maintenance and collaboration. Existing studies can be broadly categorized into retrieval-based, supervised learning-based, and LLM-based approaches.

Retrieval-based approaches generate commit messages by reusing historical code changes and their associated messages~\cite{liu2018neural, hoang2020cc2vec}. 
For example, Liu et al.~\cite{liu2018neural} introduced NNGen, which directly reuses the commit message of the nearest historical code diff, while Hoang et al.~\cite{hoang2020cc2vec} developed CC2Vec to retrieve similar commits through learned change representations. Although simple and interpretable, these approaches rely heavily on the availability of similar historical changes.

Supervised learning-based approaches formulate CMG as a code-to-text generation task using large-scale paired corpora~\cite{jiang2017automatically, liu2019generating, dong2022fira}. 
Jiang et al.~\cite{jiang2017automatically} presented CommitGen based on neural machine translation, Liu et al.~\cite{liu2019generating} enhanced generation with a pointer-generator network~\cite{see2017get}, and Dong et al.~\cite{dong2022fira} leveraged edit graphs together with graph neural networks and a Transformer decoder to better model structural code changes. 
Compared with retrieval-based approaches, these approaches better capture the semantic relationship between code changes and commit messages, but depend heavily on large-scale supervised training data.

Recent advances in LLMs have shifted CMG from sequence learning toward reasoning and retrieval-enhanced generation~\cite{li2024only, xue2024automated, xiong2026coracmg, zhang2024automatic, wu2025empirical}. 
Li et al.~\cite{li2024only} formulated CMG as a reasoning task using ReAct prompting~\cite{yao2023react}, Xue et al.~\cite{xue2024automated} introduced ERICommiter with retrieval-based in-context learning, and Xiong et al.~\cite{xiong2026coracmg} presented CoRaCMG by augmenting prompts with retrieved diff--message pairs. These studies demonstrate the strong capability of LLMs in understanding code changes and generating more informative commit messages.

Despite the rapid progress of CMG techniques, existing approaches generally treat CMG as a standalone generation task, with improvements primarily demonstrated through higher lexical and semantic similarity to developer-written references. This static process does not explicitly consider how generated messages serve software engineering activities or use the information revealed during their practical use to guide further improvement. Consequently, the practical usefulness of generated commit messages across software engineering activities remains insufficiently explored.

\subsection{Issue-Commit Traceability}
Issue--commit traceability aims to recover the code commits associated with a given issue report, thereby supporting software maintenance, debugging, and project evolution. Existing approaches can be broadly categorized into traditional information retrieval-based and heuristic-based approaches, learning-based approaches, and LLM-based approaches.

Traditional approaches mainly rely on textual similarity, handcrafted heuristics, and process information to recover issue--commit links~\cite{bachmann2010missing,wu2011relink,nguyen2012multi,schermann2015discovering,bird2009fair}. For example, Wu et al.~\cite{wu2011relink} developed ReLink by automatically learning feature-based linking criteria from historical trace links, while Nguyen et al.~\cite{nguyen2012multi} introduced MLink, which combines textual and source code features to recover links beyond textual matching. These approaches are lightweight but rely heavily on manually designed heuristics and often struggle to generalize across projects.

Learning-based approaches formulate issue--commit traceability as a supervised learning problem by learning semantic representations of issue reports and commits from labeled issue--commit pairs~\cite{ruan2019deeplink, mazrae2021hybridlinker, lan2023btlink, deng2024promptlink, wang2025mplinker, zhang2023ealink}. 
Ruan et al.~\cite{ruan2019deeplink} presented DeepLink based on recurrent neural networks~\cite{mikolov2010recurrent}, Mazrae et al.~\cite{mazrae2021hybridlinker} developed Hybrid-Linker by combining textual and non-textual features, and Lan et al.~\cite{lan2023btlink} leveraged dual BERT encoders~\cite{devlin2019bert} in BTLink to jointly encode issue text, commit code, and commit messages. Deng et al.~\cite{deng2024promptlink} and Wang et al.~\cite{wang2025mplinker} further reformulated the task as prompt-based cloze prediction, while Zhang et al.~\cite{zhang2023ealink} introduced EALink, a lightweight ranking framework based on knowledge distillation and multi-task contrastive learning. Despite their promising performance, these approaches are typically evaluated using relatively small candidate sets and therefore do not fully reflect the realistic candidate distributions encountered in large software repositories.

Recent studies have explored LLMs to further improve issue--commit traceability under realistic retrieval settings. 
Huang et al.~\cite{Huang2026Back} introduced the Realistic Distribution Setting (RDS) and developed EasyLink, a training-free retrieve-and-rerank framework based on dense retrieval and LLM reranking. More recently, Akhavan et al.~\cite{Akhavan2026LinkAnchor} proposed LinkAnchor, an agentic framework that formulates issue--commit traceability as a dynamic repository exploration task.

Unlike these studies, which focus on improving issue--commit traceability itself, our work uses traceability as a representative software maintenance activity to examine and improve the practical usefulness of generated commit messages. By incorporating traceability reasoning and feedback into CMG, our approach enables generated messages to better convey the problem--fix relationship between issue reports and code changes.

\section{Conclusion}\label{sec:conclusion}

In this study, we treat CMG as a dynamic process guided by its practical usefulness in a particular software engineering activity and propose \toolname, a collaborative multi-agent framework that integrates message generation, issue--commit traceability, and feedback-guided refinement. By leveraging traceability reasoning feedback, \toolname generates commit messages that better capture both code changes and development intent.
Experiments on six open-source projects demonstrate the effectiveness of \toolname. It achieves an average Hit@1 of 0.893 and MRR of 0.911, outperforming the state-of-the-art baselines by 13.16\% and 38.75\%, respectively.
In the future, we plan to extend this dynamic process to other software engineering activities within AI-assisted workflows.


\bibliographystyle{IEEEtran}
\balance
\bibliography{mybibfile}

@inproceedings{sun-etal-2023-chatgpt,
    title = "Is {C}hat{GPT} Good at Search? Investigating Large Language Models as Re-Ranking Agents",
    author = "Sun, Weiwei  and
      Yan, Lingyong  and
      Ma, Xinyu  and
      Wang, Shuaiqiang  and
      Ren, Pengjie  and
      Chen, Zhumin  and
      Yin, Dawei  and
      Ren, Zhaochun",
    editor = "Bouamor, Houda  and
      Pino, Juan  and
      Bali, Kalika",
    booktitle = "Proceedings of the 2023 Conference on Empirical Methods in Natural Language Processing",
    month = dec,
    year = "2023",
    address = "Singapore",
    publisher = "Association for Computational Linguistics",
    pages = "14918--14937"
}

@INPROCEEDINGS{zhang2023ealink,
  author={Zhang, Chenyuan and Wang, Yanlin and Wei, Zhao and Xu, Yong and Wang, Juhong and Li, Hui and Ji, Rongrong},
  booktitle={2023 38th IEEE/ACM International Conference on Automated Software Engineering}, 
  title={EALink: An Efficient and Accurate Pre-Trained Framework for Issue-Commit Link Recovery}, 
  year={2023},
  volume={},
  number={},
  pages={217-229},
}

@inproceedings{Huang2026Back,
  author = {Huang, Huihui and Widyasari, Ratnadira and Zhang, Ting and Irsan, Ivana Clairine and Shi, Jieke and Ang, Han Wei and Liauw, Frank and Ouh, Eng Lieh and Shar, Lwin Khin and Kang, Hong Jin and others},
  title = {Back to the Basics: Rethinking Issue-Commit Linking with LLM-Assisted Retrieval},
  booktitle = {Proceedings of the 48th IEEE/ACM International Conference on Software Engineering},
  series = {ICSE '26},
  year = {2026},
  address = {Rio de Janeiro, Brazil},
  publisher = {ACM},
  pages = {13},
}

@article{Akhavan2026LinkAnchor,
  author = {Akhavan, Arshia and Hoseinpour, Alireza and Heydarnoori, Abbas and Bagheri, Hamid and Keshani, Mehdi},
  title = {LinkAnchor: An Autonomous LLM-Based Agent for Issue-to-Commit Link Recovery},
  journal = {Proc. ACM Softw. Eng.},
  volume = {3},
  number = {FSE},
  articleno = {184},
  year = {2026},
  month = jul,
  publisher = {ACM},
  numpages = {24},
}

@inproceedings{liu2018neural,
author = {Liu, Zhongxin and Xia, Xin and Hassan, Ahmed E. and Lo, David and Xing, Zhenchang and Wang, Xinyu},
title = {Neural-machine-translation-based commit message generation: how far are we?},
year = {2018},
isbn = {9781450359375},
publisher = {Association for Computing Machinery},
address = {New York, NY, USA},
booktitle = {Proceedings of the 33rd ACM/IEEE International Conference on Automated Software Engineering},
pages = {373–384},
numpages = {12},
location = {Montpellier, France},
series = {ASE '18}
}

@inproceedings{hoang2020cc2vec,
author = {Hoang, Thong and Kang, Hong Jin and Lo, David and Lawall, Julia},
title = {CC2Vec: distributed representations of code changes},
year = {2020},
isbn = {9781450371216},
publisher = {Association for Computing Machinery},
address = {New York, NY, USA},
booktitle = {Proceedings of the ACM/IEEE 42nd International Conference on Software Engineering},
pages = {518–529},
numpages = {12},
location = {Seoul, South Korea},
series = {ICSE '20}
}

@INPROCEEDINGS{jiang2017automatically,
  author={Jiang, Siyuan and Armaly, Ameer and McMillan, Collin},
  booktitle={2017 32nd IEEE/ACM International Conference on Automated Software Engineering}, 
  title={Automatically generating commit messages from diffs using neural machine translation}, 
  year={2017},
  volume={},
  number={},
  pages={135-146},
}

@INPROCEEDINGS{liu2019generating,
  author={Liu, Qin and Liu, Zihe and Zhu, Hongming and Fan, Hongfei and Du, Bowen and Qian, Yu},
  booktitle={2019 IEEE/ACM 16th International Conference on Mining Software Repositories}, 
  title={Generating Commit Messages from Diffs using Pointer-Generator Network}, 
  year={2019},
  volume={},
  number={},
  pages={299-309},
}

@inproceedings{dong2022fira,
author = {Dong, Jinhao and Lou, Yiling and Zhu, Qihao and Sun, Zeyu and Li, Zhilin and Zhang, Wenjie and Hao, Dan},
title = {FIRA: fine-grained graph-based code change representation for automated commit message generation},
year = {2022},
isbn = {9781450392211},
publisher = {Association for Computing Machinery},
address = {New York, NY, USA},
booktitle = {Proceedings of the 44th International Conference on Software Engineering},
pages = {970–981},
numpages = {12},
location = {Pittsburgh, Pennsylvania},
series = {ICSE '22}
}

@article{li2024only,
author = {Li, Jiawei and Farag\'{o}, David and Petrov, Christian and Ahmed, Iftekhar},
title = {Only diff Is Not Enough: Generating Commit Messages Leveraging Reasoning and Action of Large Language Model},
year = {2024},
issue_date = {July 2024},
publisher = {Association for Computing Machinery},
address = {New York, NY, USA},
volume = {1},
number = {FSE},
journal = {Proc. ACM Softw. Eng.},
month = jul,
articleno = {34},
numpages = {22},
}

@article{xiong2026coracmg,
title = {CoRaCMG: Contextual retrieval-augmented framework for commit message generation},
journal = {Information and Software Technology},
volume = {196},
pages = {108169},
year = {2026},
issn = {0950-5849},
author = {Bo Xiong and Linghao Zhang and Zongen Ren and Chong Wang and Peng Liang},
}

@ARTICLE{xue2024automated,
  author={Xue, Pengyu and Wu, Linhao and Yu, Zhongxing and Jin, Zhi and Yang, Zhen and Li, Xinyi and Yang, Zhenyu and Tan, Yue},
  journal={IEEE Transactions on Software Engineering}, 
  title={Automated Commit Message Generation With Large Language Models: An Empirical Study and Beyond}, 
  year={2024},
  volume={50},
  number={12},
  pages={3208-3224},
}

@article{ruan2019deeplink,
title = {DeepLink: Recovering issue-commit links based on deep learning},
journal = {Journal of Systems and Software},
volume = {158},
pages = {110406},
year = {2019},
issn = {0164-1212},
author = {Hang Ruan and Bihuan Chen and Xin Peng and Wenyun Zhao},
}

@article{lan2023btlink,
  title={BTLink: automatic link recovery between issues and commits based on pre-trained BERT model},
  author={Lan, Jinpeng and Gong, Lina and Zhang, Jingxuan and Zhang, Haoxiang},
  journal={Empirical Software Engineering},
  volume={28},
  number={4},
  pages={103},
  year={2023},
  publisher={Springer},
}

@INPROCEEDINGS{mazrae2021hybridlinker,
  author={Mazrae, Pooya Rostami and Izadi, Maliheh and Heydarnoori, Abbas},
  booktitle={2021 IEEE International Conference on Software Maintenance and Evolution}, 
  title={Automated Recovery of Issue-Commit Links Leveraging Both Textual and Non-textual Data}, 
  year={2021},
  volume={},
  number={},
  pages={263-273},
}

@inproceedings{deng2024promptlink,
author = {Deng, Yang and Wang, Bangchao and Zou, Zhiyuan and Ye, Luyao},
title = {PromptLink: Multi-template prompt learning with adversarial training for issue-commit link recovery},
year = {2024},
isbn = {9798400710476},
publisher = {Association for Computing Machinery},
address = {New York, NY, USA},
booktitle = {Proceedings of the 18th ACM/IEEE International Symposium on Empirical Software Engineering and Measurement},
pages = {461–467},
numpages = {7},
location = {Barcelona, Spain},
series = {ESEM '24}
}

@article{wang2025mplinker,
title = {MPLinker: Multi-template Prompt-tuning with adversarial training for Issue–commit Link recovery},
journal = {Journal of Systems and Software},
volume = {223},
pages = {112351},
year = {2025},
issn = {0164-1212},
author = {Bangchao Wang and Yang Deng and Ruiqi Luo and Peng Liang and Tingting Bi},
}

@inproceedings{reimers2019sentence,
    title = "Sentence-{BERT}: Sentence Embeddings using {S}iamese {BERT}-Networks",
    author = "Reimers, Nils  and
      Gurevych, Iryna",
    editor = "Inui, Kentaro  and
      Jiang, Jing  and
      Ng, Vincent  and
      Wan, Xiaojun",
    booktitle = "Proceedings of the 2019 Conference on Empirical Methods in Natural Language Processing and the 9th International Joint Conference on Natural Language Processing",
    month = nov,
    year = "2019",
    address = "Hong Kong, China",
    publisher = "Association for Computational Linguistics",
    pages = "3982--3992",
}

@ARTICLE{douze2025faiss,
  author={Douze, Matthijs and Guzhva, Alexandr and Deng, Chengqi and Johnson, Jeff and Szilvasy, Gergely and Mazaré, Pierre-Emmanuel and Lomeli, Maria and Hosseini, Lucas and Jégou, Hervé},
  journal={IEEE Transactions on Big Data}, 
  title={The Faiss Library}, 
  year={2026},
  volume={12},
  number={2},
  pages={346-361},
}

@INPROCEEDINGS{lin2021traceability,
  author={Lin, Jinfeng and Liu, Yalin and Zeng, Qingkai and Jiang, Meng and Cleland-Huang, Jane},
  booktitle={2021 IEEE/ACM 43rd International Conference on Software Engineering}, 
  title={Traceability Transformed: Generating More Accurate Links with Pre-Trained BERT Models}, 
  year={2021},
  volume={},
  number={},
  pages={324-335},
}

@inproceedings{devlin2019bert,
    title = "{BERT}: Pre-training of Deep Bidirectional Transformers for Language Understanding",
    author = "Devlin, Jacob  and
      Chang, Ming-Wei  and
      Lee, Kenton  and
      Toutanova, Kristina",
    editor = "Burstein, Jill  and
      Doran, Christy  and
      Solorio, Thamar",
    booktitle = "Proceedings of the 2019 Conference of the North {A}merican Chapter of the Association for Computational Linguistics: Human Language Technologies, Volume 1 (Long and Short Papers)",
    month = jun,
    year = "2019",
    address = "Minneapolis, Minnesota",
    publisher = "Association for Computational Linguistics",
    pages = "4171--4186",
}

@inproceedings{dong2022semi,
author = {Dong, Liming and Zhang, He and Liu, Wei and Weng, Zhiluo and Kuang, Hongyu},
title = {Semi-supervised pre-processing for learning-based traceability framework on real-world software projects},
year = {2022},
isbn = {9781450394130},
publisher = {Association for Computing Machinery},
address = {New York, NY, USA},
booktitle = {Proceedings of the 30th ACM Joint European Software Engineering Conference and Symposium on the Foundations of Software Engineering},
pages = {570–582},
numpages = {13},
location = {Singapore, Singapore},
series = {ESEC/FSE 2022}
}

@inproceedings{radlinski2010comparing,
author = {Radlinski, Filip and Craswell, Nick},
title = {Comparing the sensitivity of information retrieval metrics},
year = {2010},
isbn = {9781450301534},
publisher = {Association for Computing Machinery},
address = {New York, NY, USA},
booktitle = {Proceedings of the 33rd International ACM SIGIR Conference on Research and Development in Information Retrieval},
pages = {667–674},
numpages = {8},
location = {Geneva, Switzerland},
series = {SIGIR '10}
}

@article{sakai2008information,
  title={On information retrieval metrics designed for evaluation with incomplete relevance assessments},
  author={Sakai, Tetsuya and Kando, Noriko},
  journal={Information Retrieval},
  volume={11},
  number={5},
  pages={447--470},
  year={2008},
  publisher={Springer},
}

@article{chmura2002kappa,
  title={Kappa coefficients in medical research},
  author={Chmura Kraemer, Helena and Periyakoil, Vyjeyanthi S and Noda, Art},
  journal={Statistics in medicine},
  volume={21},
  number={14},
  pages={2109--2129},
  year={2002},
  publisher={Wiley Online Library},
}

@book{wilcoxon1963critical,
  title={Critical values and probability levels for the Wilcoxon rank sum test and the Wilcoxon signed rank test},
  author={Wilcoxon, Frank and Katti, SK and Wilcox, Roberta A and others},
  volume={1},
  year={1963},
  publisher={American Cyanamid Pearl River, NY}
}

@article{macbeth2011cliff,
  title={Cliff's Delta Calculator: A non-parametric effect size program for two groups of observations},
  author={Macbeth, Guillermo and Razumiejczyk, Eugenia and Ledesma, Rub{\'e}n Daniel},
  journal={Universitas Psychologica},
  volume={10},
  number={2},
  pages={545--555},
  year={2011},
  publisher={Pontificia Universidad Javeriana}
}

@inproceedings{tian2022makes,
author = {Tian, Yingchen and Zhang, Yuxia and Stol, Klaas-Jan and Jiang, Lin and Liu, Hui},
title = {What makes a good commit message?},
year = {2022},
isbn = {9781450392211},
publisher = {Association for Computing Machinery},
address = {New York, NY, USA},
booktitle = {Proceedings of the 44th International Conference on Software Engineering},
pages = {2389–2401},
numpages = {13},
location = {Pittsburgh, Pennsylvania},
series = {ICSE '22}
}

@INPROCEEDINGS{li2023commit,
  author={Li, Jiawei and Ahmed, Iftekhar},
  booktitle={2023 IEEE/ACM 45th International Conference on Software Engineering}, 
  title={Commit Message Matters: Investigating Impact and Evolution of Commit Message Quality}, 
  year={2023},
  volume={},
  number={},
  pages={806-817},
}

@ARTICLE{zhang2024automatic,
  author={Zhang, Yuxia and Qiu, Zhiqing and Stol, Klaas-Jan and Zhu, Wenhui and Zhu, Jiaxin and Tian, Yingchen and Liu, Hui},
  journal={IEEE Transactions on Software Engineering}, 
  title={Automatic Commit Message Generation: A Critical Review and Directions for Future Work}, 
  year={2024},
  volume={50},
  number={4},
  pages={816-835},
}

@INPROCEEDINGS{wu2025empirical,
  author={Wu, Yifan and Wang, Yunpeng and Li, Ying and Tao, Wei and Yu, Siyu and Yang, Haowen and Jiang, Wei and Li, Jianguo},
  booktitle={2025 IEEE/ACM 47th International Conference on Software Engineering}, 
  title={An Empirical Study on Commit Message Generation Using LLMs via In-Context Learning}, 
  year={2025},
  volume={},
  number={},
  pages={553-565},
  }

@inproceedings{see2017get,
    title = "Get To The Point: Summarization with Pointer-Generator Networks",
    author = "See, Abigail  and
      Liu, Peter J.  and
      Manning, Christopher D.",
    editor = "Barzilay, Regina  and
      Kan, Min-Yen",
    booktitle = "Proceedings of the 55th Annual Meeting of the Association for Computational Linguistics (Volume 1: Long Papers)",
    month = jul,
    year = "2017",
    address = "Vancouver, Canada",
    publisher = "Association for Computational Linguistics",
    pages = "1073--1083",
}

@inproceedings{yao2023react,
  title={ReAct: Synergizing Reasoning and Acting in Language Models},
  author={Yao, Shunyu and Zhao, Jeffrey and Yu, Dian and Du, Nan and Shafran, Izhak and Narasimhan, Karthik and Cao, Yuan},
  booktitle={International Conference on Learning Representations},
  year={2023},
}

@inproceedings{bachmann2010missing,
author = {Bachmann, Adrian and Bird, Christian and Rahman, Foyzur and Devanbu, Premkumar and Bernstein, Abraham},
title = {The missing links: bugs and bug-fix commits},
year = {2010},
isbn = {9781605587912},
publisher = {Association for Computing Machinery},
address = {New York, NY, USA},
booktitle = {Proceedings of the Eighteenth ACM SIGSOFT International Symposium on Foundations of Software Engineering},
pages = {97–106},
numpages = {10},
location = {Santa Fe, New Mexico, USA},
series = {FSE '10}
}

@inproceedings{wu2011relink,
author = {Wu, Rongxin and Zhang, Hongyu and Kim, Sunghun and Cheung, Shing-Chi},
title = {ReLink: recovering links between bugs and changes},
year = {2011},
isbn = {9781450304436},
publisher = {Association for Computing Machinery},
address = {New York, NY, USA},
booktitle = {Proceedings of the 19th ACM SIGSOFT Symposium and the 13th European Conference on Foundations of Software Engineering},
pages = {15–25},
numpages = {11},
location = {Szeged, Hungary},
series = {ESEC/FSE '11}
}

@inproceedings{nguyen2012multi,
author = {Nguyen, Anh Tuan and Nguyen, Tung Thanh and Nguyen, Hoan Anh and Nguyen, Tien N.},
title = {Multi-layered approach for recovering links between bug reports and fixes},
year = {2012},
isbn = {9781450316149},
publisher = {Association for Computing Machinery},
address = {New York, NY, USA},
booktitle = {Proceedings of the ACM SIGSOFT 20th International Symposium on the Foundations of Software Engineering},
articleno = {63},
numpages = {11},
location = {Cary, North Carolina},
series = {FSE '12}
}

@INPROCEEDINGS{schermann2015discovering,
  author={Schermann, Gerald and Brandtner, Martin and Panichella, Sebastiano and Leitner, Philipp and Gall, Harald},
  booktitle={2015 IEEE 23rd International Conference on Program Comprehension}, 
  title={Discovering Loners and Phantoms in Commit and Issue Data}, 
  year={2015},
  volume={},
  number={},
  pages={4-14},
  }

@inproceedings{bird2009fair,
author = {Bird, Christian and Bachmann, Adrian and Aune, Eirik and Duffy, John and Bernstein, Abraham and Filkov, Vladimir and Devanbu, Premkumar},
title = {Fair and balanced? bias in bug-fix datasets},
year = {2009},
isbn = {9781605580012},
publisher = {Association for Computing Machinery},
address = {New York, NY, USA},
booktitle = {Proceedings of the 7th Joint Meeting of the European Software Engineering Conference and the ACM SIGSOFT Symposium on The Foundations of Software Engineering},
pages = {121–130},
numpages = {10},
location = {Amsterdam, The Netherlands},
series = {ESEC/FSE '09}
}

@inproceedings{mikolov2010recurrent,
  title={Recurrent neural network based language model.},
  author={Mikolov, Tomas and Karafi{\'a}t, Martin and Burget, Lukas and Cernock{\`y}, Jan and Khudanpur, Sanjeev},
  booktitle={Interspeech},
  volume={2},
  number={3},
  pages={1045--1048},
  year={2010},
  organization={Makuhari},
}

@inproceedings{zheng2023judging,
 author = {Zheng, Lianmin and Chiang, Wei-Lin and Sheng, Ying and Zhuang, Siyuan and Wu, Zhanghao and Zhuang, Yonghao and Lin, Zi and Li, Zhuohan and Li, Dacheng and Xing, Eric and Zhang, Hao and Gonzalez, Joseph and Stoica, Ion},
 booktitle = {Advances in Neural Information Processing Systems},
 editor = {A. Oh and T. Naumann and A. Globerson and K. Saenko and M. Hardt and S. Levine},
 pages = {46595--46623},
 publisher = {Curran Associates, Inc.},
 title = {Judging LLM-as-a-Judge with MT-Bench and Chatbot Arena},
 volume = {36},
 year = {2023}
}

@inproceedings{kim2024prometheus,
 author = {Kim, Seungone and Shin, Jay and cho, yejin and Jang, Joel and Longpre, Shayne and Lee, Hwaran and Yun, Sangdoo and Shin, Ryan, S and Kim, Sungdong and Thorne, James and Seo, Minjoon},
 booktitle = {International Conference on Learning Representations},
 pages = {29927--29962},
 title = {Prometheus: Inducing Fine-Grained Evaluation Capability in Language Models},
 volume = {2024},
 year = {2024}
}

@inproceedings{liu2023g,
    title = "{G}-Eval: {NLG} Evaluation using Gpt-4 with Better Human Alignment",
    author = "Liu, Yang  and
      Iter, Dan  and
      Xu, Yichong  and
      Wang, Shuohang  and
      Xu, Ruochen  and
      Zhu, Chenguang",
    editor = "Bouamor, Houda  and
      Pino, Juan  and
      Bali, Kalika",
    booktitle = "Proceedings of the 2023 Conference on Empirical Methods in Natural Language Processing",
    month = dec,
    year = "2023",
    address = "Singapore",
    publisher = "Association for Computational Linguistics",
    pages = "2511--2522",
}

@article{badampudi2023modern,
author = {Badampudi, Deepika and Unterkalmsteiner, Michael and Britto, Ricardo},
title = {Modern Code Reviews—Survey of Literature and Practice},
year = {2023},
issue_date = {July 2023},
publisher = {Association for Computing Machinery},
address = {New York, NY, USA},
volume = {32},
number = {4},
issn = {1049-331X},
journal = {ACM Trans. Softw. Eng. Methodol.},
month = may,
articleno = {107},
numpages = {61},
}

@inproceedings {fang2024large,
author = {Chongzhou Fang and Ning Miao and Shaurya Srivastav and Jialin Liu and Ruoyu Zhang and Ruijie Fang and Asmita and Ryan Tsang and Najmeh Nazari and Han Wang and Houman Homayoun},
title = {Large Language Models for Code Analysis: Do {LLMs} Really Do Their Job?},
booktitle = {33rd USENIX Security Symposium (USENIX Security 24)},
year = {2024},
isbn = {978-1-939133-44-1},
address = {Philadelphia, PA},
pages = {829--846},
publisher = {USENIX Association},
month = aug
}

@INPROCEEDINGS{ren2023deeparc,
  author={Ren, Xiaoning and Lin, Yun and Xue, Yinxing and Liu, Ruofan and Sun, Jun and Feng, Zhiyong and Dong, Jin Song},
  booktitle={2023 IEEE/ACM 45th International Conference on Software Engineering}, 
  title={DeepArc: Modularizing Neural Networks for the Model Maintenance}, 
  year={2023},
  volume={},
  number={},
  pages={1008-1019},
}

@article{zhang2026survey,
  title={A survey on large language models for software engineering},
  author={Zhang, Quanjun and Fang, Chunrong and Xie, Yang and Zhang, Yaxin and Yu, Shengcheng and Sun, Weisong and Yang, Yun and Chen, Zhenyu},
  journal={Sci. China Inf. Sci.},
  volume={69},
  number={4},
  pages={141102},
  year={2026},
  publisher={Springer},
}

@inproceedings{banerjee2005meteor,
  title={METEOR: An automatic metric for MT evaluation with improved correlation with human judgments},
  author={Banerjee, Satanjeev and Lavie, Alon},
  booktitle={Proceedings of the acl workshop on intrinsic and extrinsic evaluation measures for machine translation and/or summarization},
  pages={65--72},
  year={2005}
}

@inproceedings{papineni2002bleu,
  title={Bleu: a method for automatic evaluation of machine translation},
  author={Papineni, Kishore and Roukos, Salim and Ward, Todd and Zhu, Wei-Jing},
  booktitle={Proceedings of the 40th annual meeting of the Association for Computational Linguistics},
  pages={311--318},
  year={2002}
}

@inproceedings{lin2004rouge,
  title={Rouge: A package for automatic evaluation of summaries},
  author={Lin, Chin-Yew},
  booktitle={Text summarization branches out},
  pages={74--81},
  year={2004}
}

@article{zhang2019bertscore,
  title={Bertscore: Evaluating text generation with bert},
  author={Zhang, Tianyi and Kishore, Varsha and Wu, Felix and Weinberger, Kilian Q and Artzi, Yoav},
  journal={arXiv preprint arXiv:1904.09675},
  year={2019}
}

\end{document}